\documentclass{aa}  
\usepackage[varg]{txfonts}
\usepackage{graphicx}
\usepackage{longtable}
\usepackage{natbib}

\usepackage[colorlinks=true, citecolor=blue, linkcolor=blue]{hyperref}

\begin{document}

   \title{Proper motion study of the 6.7~GHz methanol maser rings.}
   \subtitle{I. A sample of sources with little variation}

   \author{A. Bartkiewicz
          \inst{1} \href{https://orcid.org/0000-0002-6466-117X}{\includegraphics[scale=0.5]{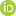}},
          A. Sanna 
          \inst{2} \href{https://orcid.org/0000-0001-7960-4912}{\includegraphics[scale=0.5]{orcid.png}},
          M. Szymczak
          \inst{1} \href{https://orcid.org/0000-0002-1482-8189}{\includegraphics[scale=0.5]{orcid.png}},
          L. Moscadelli
          \inst{3} \href{https://orcid.org/0000-0002-8517-8881}{\includegraphics[scale=0.5]{orcid.png}},
          H.J. van Langevelde
          \inst{4,5,6} \href{https://orcid.org/0000-0002-0230-5946}{\includegraphics[scale=0.5]{orcid.png}},
          P. Wolak
          \inst{1} \href{https://orcid.org/0000-0002-5413-2573}{\includegraphics[scale=0.5]{orcid.png}},
          A. Kobak
          \inst{1} \href{https://orcid.org/0000-0002-1206-9887}{\includegraphics[scale=0.5]{orcid.png}}
          \and
          M. Durjasz
          \inst{1} \href{https://orcid.org/0000-0001-7952-0305}{\includegraphics[scale=0.5]{orcid.png}}
          }

   \institute{Institute of Astronomy, Faculty of Physics, Astronomy and Informatics, Nicolaus Copernicus University, Grudziadzka 5, 87-100 Torun, Poland,
\email{annan@astro.umk.pl}
         \and
         INAF, Osservatorio Astronomico di Cagliari, via della Scienza 5, 09047, Selargius, Italy
         \and
         INAF, Osservatorio Astrofisico di Arcetri, Largo E. Fermi 5, 50125, Firenze, Italy
         \and
             Joint Institute for VLBI ERIC (JIVE), Oude Hoogeveensedijk 4, 7991 PD Dwingeloo, The
Netherlands
         \and
             Sterrewacht Leiden, Leiden University, Postbus 9513, 2300 RA Leiden, The Netherlands
        \and 
            University of New Mexico, Department of Physics and Astronomy, 
Albuquerque, NM 87131, USA
             }

   \date{Received 5 February 2024; accepted 28 March 2024}

 
  \abstract
   {Methanol masers at 6.7~GHz are well-known signposts of high-mass star-forming regions. Due to their high brightness, they enable us to derive the three-dimensional gas kinematics near protostars and young stars.}
   {We aim to understand the origin of the ring-like structures outlined by methanol maser emission in a number of sources. This emission
   could be, a priori, spatially associated with an outflow and/or disc around a high-mass protostar. In cases of expansion or rotation, maser proper motions should be, for instance, diverging from the ring centre or perpendicular to the ring radius.} 
   {Using sensitive, three-epoch observations spanning over eight years with the European VLBI Network, we have started the most direct investigations of maser rings using very accurate proper motion measurements with uncertainties below $\sim$1\,km~s$^{-1}$.}
   {We present full results for the five targets of our sample, G23.207$-$00.377, G23.389$+$00.185, G28.817$+$00.365, G31.047$+$00.356, and G31.581$+$00.077, where proper motions show similar characteristics; maser cloudlets do not move inwards towards the centre of the rings but rather outwards. We also include the most circular source, G23.657$-$00.127, in the discussion as a reference. 
   The magnitude of maser proper motions ranges from a maximum of about 13\,km~s$^{-1}$ to 0.5~km~s$^{-1}$. In two of the five sources with a high number of maser spots ($>$\,100), namely G23.207$-$00.377 and  G23.389$+$00.185, we show that the size of the best elliptical model, fitted to the distribution of persistent masers, increases in time in a manner similar to the case of G23.657$-$00.127. Moreover, we checked the separations between the pairs of spots from distinct regions, and we were able to assess that G28.817$+$00.365 and G31.047$+$00.356 can be interpreted as showing expanding motions. We analysed the profiles of single maser cloudlets and studied their variability. Contrary to single-dish studies, the interferometric data indicate variability of the emission of single-masing cloudlets.}
  {In five of the six targets, namely, G23.207$-$00.377, G23.389$+$00.185, G23.657$-$00.127, G28.817$+$00.365, and G31.047$+$00.356, expansion motions prevail. Only in the case of G31.581$+$00.077 can a scenario of disc-like rotation not be excluded. Complementary observations of thermal tracers as well as searching for ultra-compact H~{\small II} regions in the same sources are needed. Although the overall morphology of the maser emission has remained stable, the intensities of individual maser cloudlets varied from epoch to epoch, suggesting internal instabilities.}


\titlerunning{EVN proper motion study of the 6.7~GHz methanol maser rings. I.}
\authorrunning{A. Bartkiewicz et al.}

   \maketitle
%

\nolinenumbers
\section{Introduction}
Young high-mass stars with masses larger than 8\,M$_{\odot}$ form inside star-forming regions, and they are still challenging targets in modern astrophysics, as they are distant (distances of a few kiloparsecs), obscured at optical and near-infrared wavelengths, and evolve on short timescales. These massive young stellar objects (MYSOs) can be studied by means of 6.7~GHz methanol maser observations \citep{m91} because this line has been found to uniquely trace their environment (e.g.~\citealt{br13} and references therein). Very long baseline interferometry (VLBI) allows us to derive the detailed spatial structure of methanol maser emission at milliarcsecond resolution (e.g. \citealt{mos11}, \citeyear{mosmas11}, \citealt{n2023}) and investigate the small (a few astronomical units) regions of neutral gas in the vicinity of MYSOs (at radii of 1000s AU). Since the first VLBI observations of the 6.7~GHz methanol maser emission almost three decades ago, maser spots have been shown to be found in diverse structures. These structures can consist of a few or many maser spots, having either linear or arched spatial distributions, which often show regular velocity gradients as well (e.g.~\citealt{n93}, \citealt{p98}, \citealt{w98}, \citealt{m00}). With the increase of the VLBI sensitivity, ring-like structures were also discovered (\citealt{b09}, \citealt{t11}, \citealt{f14}). However, the general question about where the methanol maser emission arises has not yet been answered. Recently, proper motion studies using VLBI data have added further information to address this question. Such observations provide the most direct measurements of the gas kinematics near MYSOs, although long time baselines are required -- typically of a few years -- to trace shifts on the order of a few milliarcseconds for a single maser spot.

A series of works by 
\cite{mos00,mos05,mos07,mos11} demonstrated the ability of water masers to trace 3D 
velocities of gas, which usually expands from MYSOs at several 10\,km\,s$^{-1}$. Towards two MYSOs,
 G16.59$-$0.05 and G23.01$-$0.41,
 \cite{s10a,s10b} also reported accurate multi-epoch proper motion measurements of the 6.7~GHz masers for the first
time, using a time baseline of two years. This maser 
emission arises close to
the central protostar (within 1000\,AU) and traces gas
either rotating and/or expanding with mean (transverse) velocities of 6--7\,km~s$^{-1}$. Similarly, \cite{mosmas11}
showed that the methanol masers in IRAS~20126$+$4104 are associated with a rotating disc while also moving in the direction of the jet axis. 
A special case was further discovered in AFGL\,5142, where the methanol masers provided a direct measurement of gas infall within a radius of
300~AU from an intermediate-mass protostar (\citealt{g11}). In Cepheus A HW2, the velocity field is dominated by an infall component of about 2\,km~s$^{-1}$ down to a radius of 300~AU, where a rotational component of 4\,km~s$^{-1}$ becomes dominant, as revealed by the 6.7~GHz methanol maser studies (\citealt{t11}, \citealt{s14}, \citealt{s17}).

The following project aims to measure the proper motions of those 6.7\,GHz methanol masers distributed along ring-like structures. We selected
twelve objects from our sample of 31 sources that were observed from 2004 to 2007 by \cite{b09} using the European VLBI
Network\footnote{The European VLBI Network is a joint facility of independent European, African, Asian, and North American radio astronomy institutes.
Scientific results from data presented in this publication are derived from the following EVN project codes: EN003, EB031, EB034, EB052.} (EVN) and repeated the observations with a similar setup in 2013 and 2015. Since 6.7\,GHz methanol masers typically trace gas velocities of less than 10\,km~s$^{-1}$
and the targets in our sample have distances within about 4\,kpc from the Sun, one would expect to detect position shifts of ~3\,mas after six~years.
These angular shifts are about half the EVN beam size and are easily detectable when there is a sufficient signal-to-noise ratio (S/N). 

In this paper, we present proper motion results for five of the 12 sources in our sample. The first sub-sample concerns targets whose ring-like morphology was stable and where we identified persistent spots distributed along the whole structure over about eight years. Single-dish monitoring showed little or no variation of the 6.7~GHz line emission towards the same targets \citep{sz18}. Another source, G23.657-00.127, showing a nearly circular morphology, also belongs to this sub-sample but was analysed and described in detail in \cite{b20}. Here, we also include some results of G23.657$-$00.127 for completeness.

\section{Observations and data reduction}
The observations of the 6668.519~MHz methanol maser line taken from 2004 to 2007 are described in \cite{b09} and are referred to
here as epoch 1 (hereafter E1). The observations for epoch 2 and epoch 3 (hereafter E2 and E3, respectively) were carried out using the EVN on March 2013 and 2015 (Table~\ref{table1}). Data were
obtained from the following antennas: Jodrell Bank, Effelsberg, Medicina, Onsala, Noto, Torun, Westerbork, and Yebes. Details of the observations, such as
pointing positions of the targets, phase calibrators, and exact dates of observations for the five masers, are given in Table~\ref{table1}. Each observing run
lasted for 9 hours. The phase-referencing mode was used, repeating a loop of 5\,min between the maser target (3.25~min) and a phase calibrator from the VLBA
Calibrator List (1.75~min). The calibrator 3C345 was observed during a few 15-min scans at each run; these scans were used to calibrate the bandpass and
instrumental phases. Three maser sources were observed in each 9-hr run; these targets were selected to be within a few degrees of each other, in projection
on the sky, and of similar maser emission velocities (Table~\ref{table1}). This process yielded the on-source time for each maser of ca.~1.5~hr. The SFXC software correlator in JIVE was used with a 2-second integration time \citep{k15}. Selection of 1024 channels in the 2~MHz bandwidth resulted in a 0.1~km~s$^{-1}$ resolution. To increase S/N on the phase-reference source, eight 2-MHz BBCs per polarisation were used for a second correlator pass with 128 channels per BBC.  

The data reduction was carried out with the Astronomical Image Processing System (AIPS) developed by the National Radio Astronomy Observatory (NRAO) with standard procedures for spectral line observations. The Effelsberg antenna was set as the reference. As we aimed for detailed internal motion studies of masers and relied on the relative motions, fringe fitting (task FRING) was performed on the strongest maser channel from the first-epoch observations \citep{b09}. This allowed us to exclude motions due to annual parallax and Galactic rotation and also provided the highest possible S/N to get the most accurate measurements of the relative positions. To obtain the positions of the methanol maser spots in all channel maps, we used the AIPS tasks JMFIT or SAD (depending on the number of spots per source) fitting two-dimensional Gaussian models to the compact maser emission. The formal fitting errors were typically less than 0.1\,mas per spot. 
The rms noise level (1$\sigma_{\rm rms}$) in line-free channels was typically 4 or 6 mJy for each source, and we searched for maser emission above a threshold of 7$\sigma_{\rm rms}$ in each channel map. The synthesised beams, obtained with natural weightings in E2, are also listed in Table~\ref{table1}. 

\begin{table*}
\centering
\caption{Details of EVN observations.}
\label{table1}
\begin{tabular}{@{}lllcccc@{}}
\hline
\multicolumn{1}{c}{Source$^{a}$} & \multicolumn{2}{c}{Pointing positions (J2000)} & Velocity & Phase-calibrator & Observing & Synthesised beam$^{c}$\\
 & & & peak & & runs$^{b}$ (and 1$\sigma_{\rm rms}$ levels)\\
\multicolumn{1}{c}{Gll.lll$\pm$bb.bbb} & \multicolumn{1}{c}{RA (h:m:s)} & \multicolumn{1}{c}{Dec ($^{\rm o}$:':'')} & \multicolumn{1}{c}{$({\rm km~s}^{-1})$} &  &  (mJy) &
\multicolumn{1}{c}{(mas$\times$mas; PA(\degr))}\\
\hline
 G23.207$-$00.377 &18:34:55.21212 &$-$08:49:11.8926 & 77.1 & J1825$-$0737 & A(4), E(4) & 8.3$\times$4.1; $-$27\\
 G23.389$+$00.185 &18:33:14.32477 &$-$08:23:57.4723 & 75.4 & J1825$-$0737 & A(4), E(4) & 8.3$\times$4.1; $-$27\\
G23.657$-$00.127 &18:34:51.56482 &$-$08:18:21.3045 & 82.6 & J1825$-$0737 & A(4), E(4) & 9.0$\times$4.0; $-$28\\
 G28.817$+$00.365 &18:42:37.34797 &$-$03:29:40.9216 & 90.7 & J1834$-$0301 & C(4), G(4) & 7.9$\times$4.1; $-$30\\
 G31.047$+$00.356 &18:46:43.85506 &$-$01:30:54.1551 & 80.7 & J1834$-$0301 & C(4), G(4) & 7.5$\times$4.1; $-$32\\
 G31.581$+$00.077 &18:48:41.94108 &$-$01:10:02.5281 & 95.6 & J1834$-$0301 & D(6), H(6) & 9.3$\times$5.5; $+$4.7\\
\hline
\end{tabular}
\tablefoot{$^a$ The names are the Galactic coordinates of the brightest maser spots as in \cite{b09}. $^b$ The project EB052 was observed in the following runs: A (2/03/2013), 
C (4/03/2013), D (5/03/2013), E (15/03/2015), 
G (18/03/2015) and H (19/03/2015). $^{c}$ The sizes as in the runs A, C, D carried out in 2013.}
\end{table*}

\subsection{Proper motions measurements}
In order to study the proper motion of the masers, we followed the general procedure (method\,1): i) First we determined the maser `cloudlets' at each epoch, where a maser cloudlet is defined as a combination of at least three (in rare cases two) maser spots appearing at
consecutive spectral channels and coinciding in position within half the synthesised beam (e.g. \citealt{s17}). ii) Next, we identified the persistent maser
cloudlets over the epochs on the basis of their linear motion with respect to a bright, compact, and spectrally stable feature. Relative proper motions were derived via linear
fits of cloudlet displacements over the three epochs. iii) Last, we subtracted the average proper motion of persistent cloudlets from the proper motion of each cloudlet. As a result of this, we defined a centre of motion approximating the MYSO's rest frame, with respect to which we calculated the proper motions. This is equivalent to calculating the geometric mean of the positions of all persistent maser cloudlets at each epoch and then referring the proper motions to this point. We have already discussed this approach in \cite{b20} for G23.657$-$00.127. 

For two targets with a significant number of spots ($>$100), namely G23.207$-$00.377 and G23.389$+$00.185 (as well as in G23.657$-$00.127 as presented in \cite{b20}), we also followed a second approach for the proper motion analysis, taking advantage of the best-fitted ellipses to the
maser distribution. This second approach (method\,2) involved four steps: i) selection of maser spots that were visible at all three epochs; ii) fitting the flux-weighted
ellipses to the selected maser spots at each epoch using the code by \cite{f99}; iii) aligning of the centres of best-fitted ellipses with
respect to the first epoch; iv) constructing the averaged proper motion vector (i.e. the displacements between pairs of spots within a group were summed and divided by the number of pairs for each group of maser spots that were clearly separated from each other). Assuming that the methanol emission traces an ellipse and that the centre of the ellipse coincides with the MYSO position enabled us to remove any bulk motion of the ring in the plane of the sky and obtain the proper motions of masers relative to MYSO. 

As mentioned earlier, the formal fitting errors of maser spot positions were typically negligible, less than 0.1~mas. This evidence strengthens the accuracy of the proper motion
measurements. Also, the long time baseline (8-11\,yr) of the observations is an advantage of the present study, considering that the
targets are at distances of a few kiloparsecs. We assumed in our studies that co-spatial maser spots at the same LSR velocity identified at different epochs trace the same volume of gas. In method 1, we derived the proper motions via linear fits of cloudlet displacements over three epochs. We drew `error cones' for each proper motion vector, as derived from the uncertainty of the linear fits in each direction (RA and Dec). In method 2, we derived the averaged proper motion vectors between E1-E2 and E1-E3 data and drew error cones as a range of these two directions.

\subsection{Single-dish monitoring}
The whole sample has been monitored 
with the 32-m Torun radio telescope under the project described 
in \cite{sz18} in the period
2009-2013. In addition, follow-up observations have been ongoing since early 2018. The spectral resolution was 0.09~km~s$^{-1}$ after Hanning smoothing, a typical sensitivity (3$\sigma$) was 0.8\,Jy, and the flux density calibration accuracy was about 10\%.

\section{Results}
In this section, we present the results for five sources that show similar
characteristics in their proper motions and for which the single-dish monitoring showed little or no variation of the 6.7~GHz line \citet{sz18}. The archetype ring-like source G23.657$-$00.127 is presented in detail in \cite{b20}, and here we only report some characteristics for comparison. 

\begin{figure}
\centering
\includegraphics[scale=0.53, trim={0.8cm 0 0 0},clip]{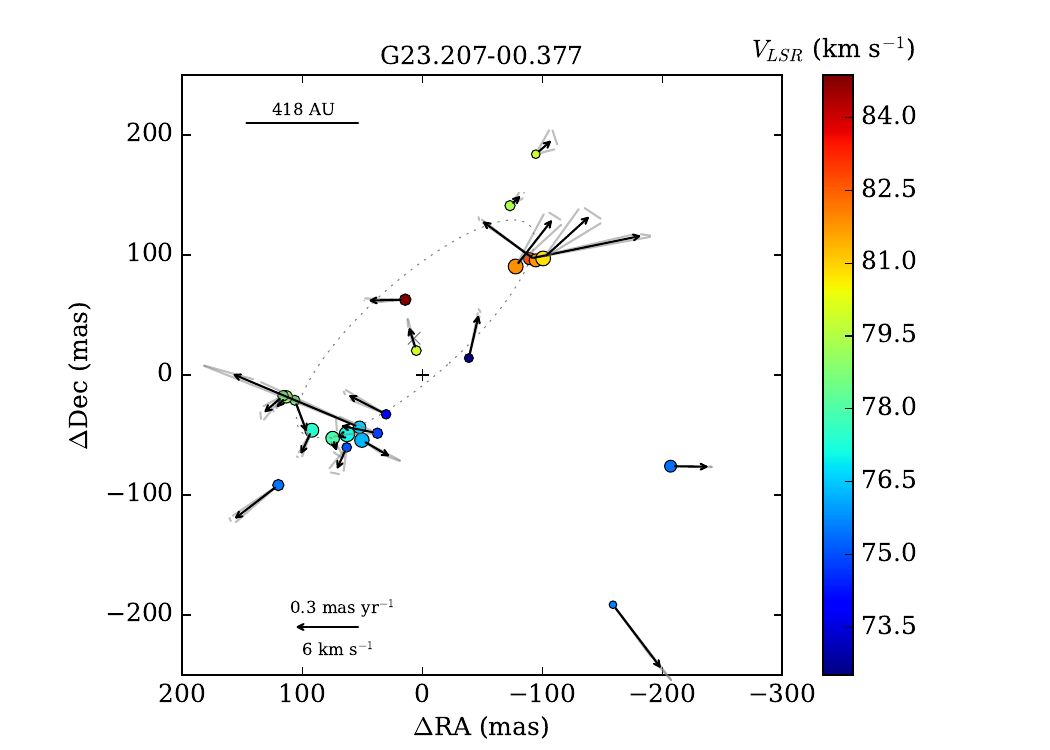}
\includegraphics[scale=0.53, trim={0.8cm 0 0 0},clip]{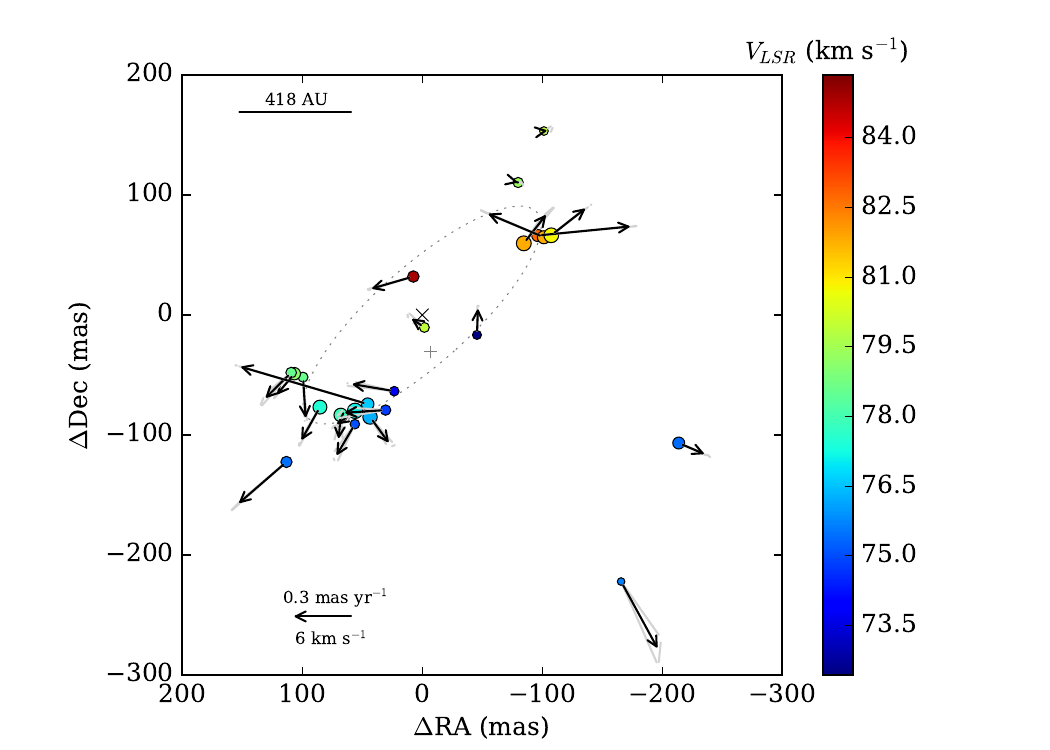}
\caption{Proper motions of 6.7~GHz methanol maser cloudlets in G23.207$-$00.377. Top: Proper motions as measured relative to the centre of motion. The black arrows indicate the best fits of the relative proper motion over the three-epoch data, and the uncertainties are marked by the grey triangles. The centre of motion is marked by the plus sign and explained in Sect.~2.1 (method 1); the (0,0) point is offset by ($-$62.59255~mas, 50.5022~mas) relative to the brightest spots listed in Table~\ref{table1}. The circle sizes are proportional to the logarithm of the flux of the brightest maser spots of each cloudlet at epoch E1, and the colours relate to their LSR velocities as shown on the vertical wedge. The dotted ellipse traces the best flux-weighted fit to all cloudlets as detected in E1 (Table~\ref{ellipses}), and its centre is marked with a cross. Bottom: Proper motions estimated using method 2 by fitting the ellipse at each epoch and aligning their centres. The (0,0) point, marked by a cross, corresponds to the centre of the best-fitted ellipse in E1 (Table~\ref{ellipses}). The black arrows represent the averaged proper motion vectors between E1-E2 and E1-E3 data.}
\label{g23207cm}  
\end{figure}

\begin{figure}
\sidecaption
\centering
\includegraphics[scale=0.53, trim={0.8cm 0 0 0},clip]{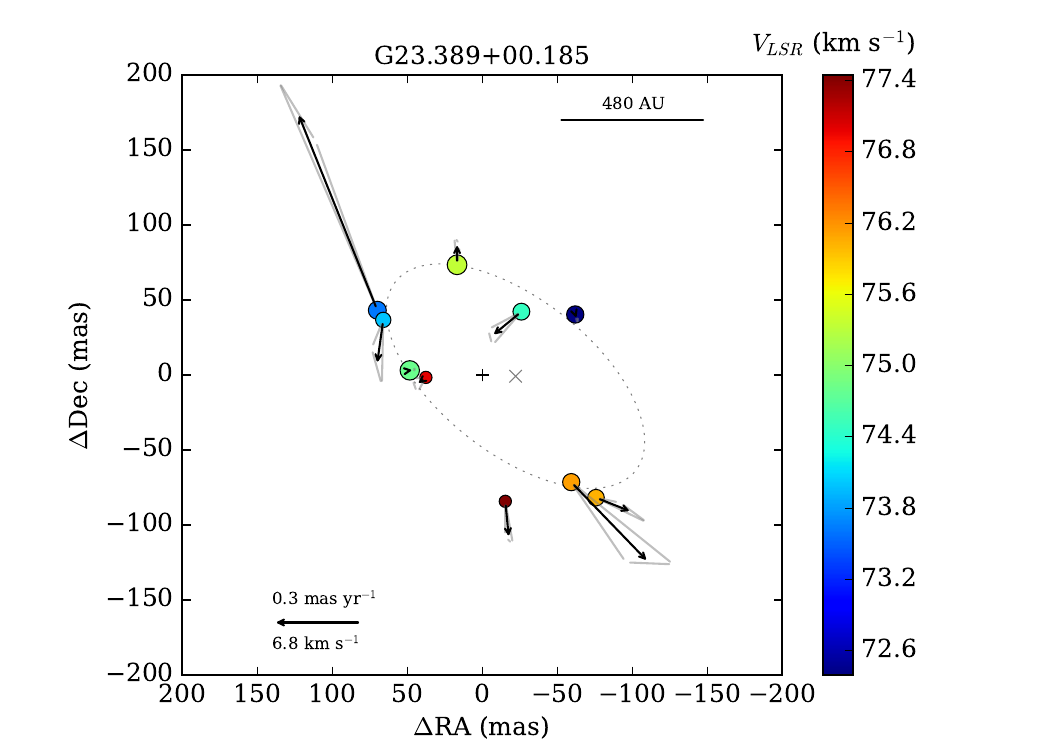}
\includegraphics[scale=0.53, trim={0.8cm 0 0 0},clip]{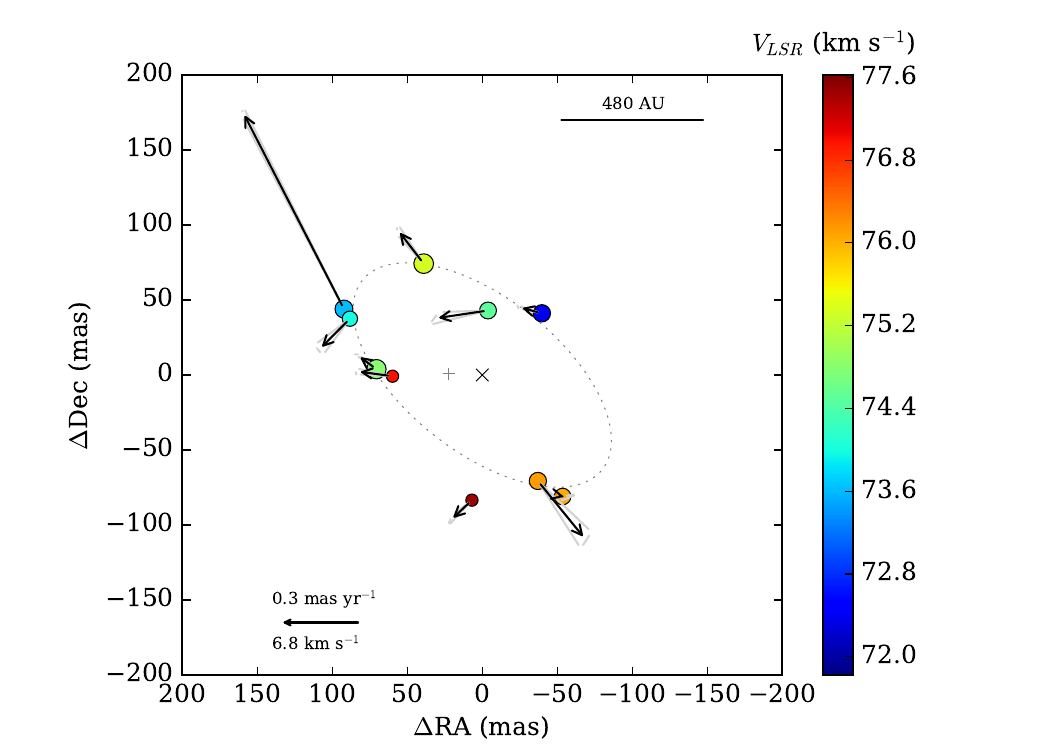}
\caption{Same as Fig.~\ref{g23207cm} but for G23.389$+$00.185. The centre of motion (marked by the plus sign) from method 1, the (0,0) point in the left panel, is shifted by ($-$16.8016~mas, $-$73.825~mas) relative to the brightest spots listed in Table~\ref{table1}. The (0,0) point in the right panel is defined by the centre of the best-fitted ellipse in E1 (Table~\ref{ellipses}).} 
\label{g23389cm}  
\end{figure}

\begin{figure}
\centering
\includegraphics[scale=0.53, trim={0.8cm 0 0 0},clip]{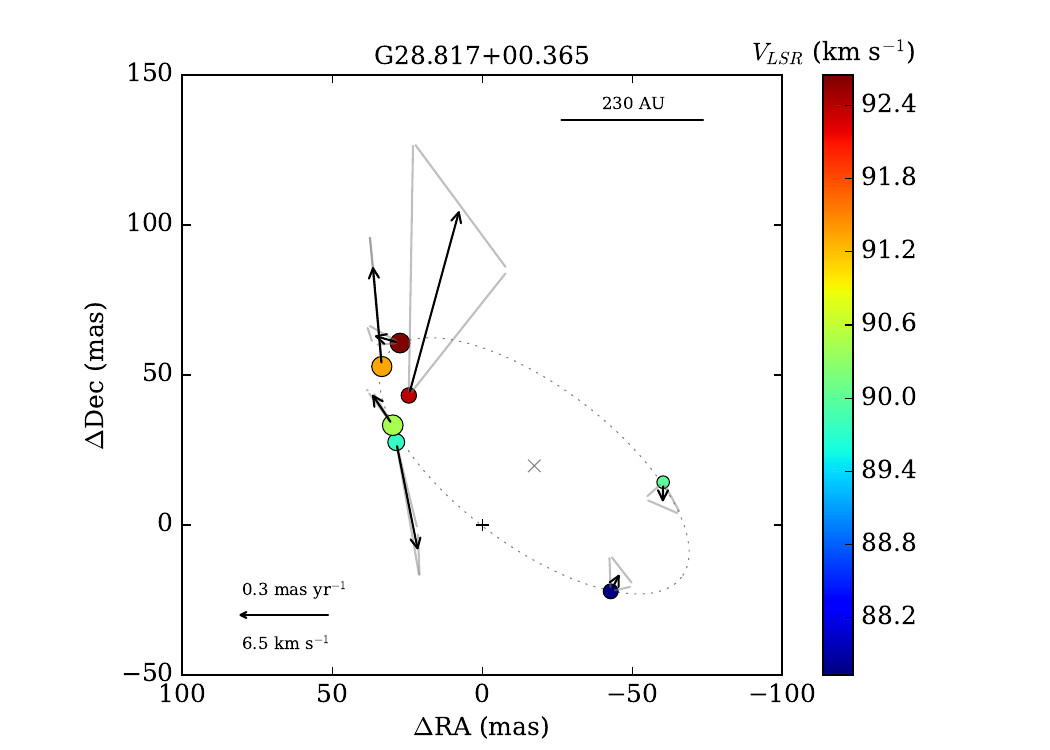}
\caption{Same as Fig.~\ref{g23207cm} but for G28.817$+$00.365. Due to the significantly smaller number of cloudlets, only method 1 was used for the proper motion studies. The centre of motion is marked by the plus sign, the (0,0) point, and it is shifted by ($-$30.0824~mas, $-$33.9852~mas) relative to the brightest spots listed in Table~\ref{table1}.} 
\label{g28817cm}  
\end{figure}

\begin{figure*}
\centering
\includegraphics[scale=0.53, trim={0.8cm 0 0 0},clip]{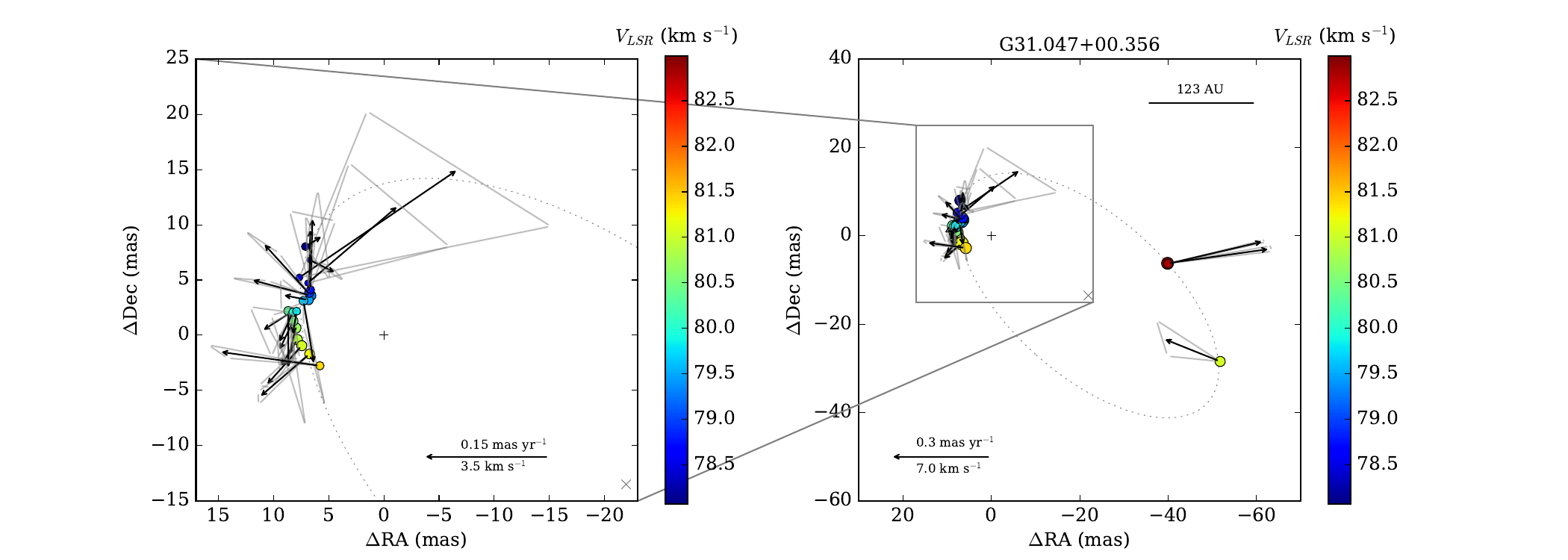}
\caption{Same as Fig.~\ref{g28817cm} but for G31.047$+$00.356. We note that 
all maser spots (instead of cloudlets, as in the other targets) are presented due to the complex structure in the western region, which is also zoomed-in for clarity in the left panel. The centre of motion is marked by the plus sign, the (0,0) point, and it is shifted by ($-$7.9729~mas, $-$0.6561~mas) relative to the brightest spots listed in Table~\ref{table1}.}
\label{g31047cm}  
\end{figure*}

\begin{figure}
\centering
\includegraphics[scale=0.6, trim={0.8cm 0 0 0},clip]{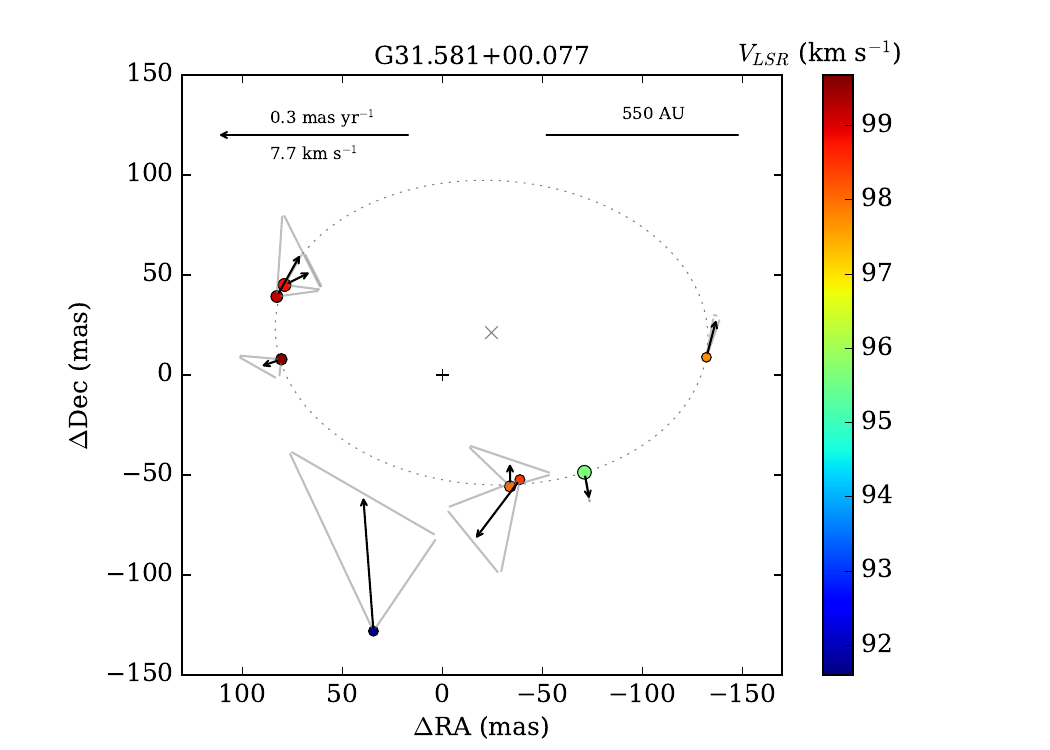}
\caption{Same as Fig.~\ref{g28817cm} but for G31.581$+$00.077. The centre of motion marked by the plus sign, the (0,0) point, is shifted by (71.0853~mas, 48.6932~mas) relative to the brightest spots listed in Table~\ref{table1}.}
\label{g31581cm}  
\end{figure}

In Fig.~\ref{threeepochs} we present the distribution of the 6.7~GHz methanol maser emission (all visible spots at single maser channel maps) at all three epochs of observations. Also, we present the spectra of the cloudlets and analyse their Gaussian profiles. The dynamic spectra of the 6.7~GHz methanol maser variability, obtained using
the Torun dish, are presented in Fig.~\ref{spectravar}. Distances to the targets, derived either from trigonometric parallax measurements or kinematic distances, have been
added to Table~\ref{ellipses}, where the flux-weighted ellipse fits to the distribution of the persisting cloudlets are summarised. The maser cloudlets with derived proper motions are listed in Table~\ref{clouds}
and presented in Figs \ref{g23207cm}-\ref{g31581cm}.  One can note that the overall structures in all cases persisted over almost eight~years (G28.817$+$00.365, G31.047$+$00.356) or even 10.5~years (G23.207$-$00.377, G23.389$+$00.185, G23.657$-$00.127). In general, displacements of the cloudlets suggest outward motions from the centres of the rings (i.e. fitted ellipses).

For G23.207$-$00.377, the following number of spots were registered during observations in 2004, 2013, and 2015: 218, 219, and 250,
respectively (Fig.~\ref{threeepochs}). We identified 140 spots that appeared in all three epochs and that can be grouped into 23 cloudlets. These cloudlets show Gaussian profiles (Fig.~\ref{threeepochs}) with varying flux densities for most of them between 2004 and 2013 (Table~\ref{clouds}).  The brightest cloudlet in 2004, at the LSR velocity of 77.1~km~s$^{-1}$, decreased by a factor of more than four after 8.5~yr. This is contrary to the results of single-dish monitoring, where the overall emission from this target 
in recent years (since 2009) shows little spectral variation (Fig.~\ref{spectravar}; see also \cite{sz18}). In the poor signal-to-noise ratio observations and without any possibility to spatially resolve spectrally blended features, it is impossible to detect single-cloudlet variability. As we stated in \cite{b20}, the internal instabilities of the masing regions may reduce the velocity coherence along the path of maser rays and induce flux variations. However, we also note that the difference in the variability patterns from VLBI and the 32-m single-dish monitoring possibly results from the blending effect, as VLBI probes the most compact cloud or the core of the cloudlet. 

We selected 20 cloudlets with approximately linear motions over three epochs (with respect to the brightest one) and used them for calculating the average proper motion. The remaining three cloudlets are marked in Table~\ref{clouds}. Figure~\ref{g23207cm} presents the derived proper motions for G23.207$-$00.377. This plot highlights the fact that the majority of cloudlets move outwards from the centroid of their distribution. Their motions range from 0.06 to 0.58~mas~yr$^{-1}$, which at a distance of 4.18~kpc \citep{r19} correspond to linear velocities from 1.2 to 11.6~km~s$^{-1}$. We calculated the radial and tangential components of the proper motions (Fig.~\ref{fig_rot_exp}), and there is a clear outward radial motion in the sky with a mean value of 2.61$\pm$0.09~km~s$^{-1}$. The tangential components do not show any consistent rotation pattern. 
Following method~2 to measure maser proper motions, by aligning the centre of the ellipses at different epochs, we obtained results consistent with method~1 after excluding the blueshifted spots to the SW of the ellipse fit. The major axis of the best-fitted ellipses increased in time, with a linear size of 1072, 1082, and 1084~AU in 2004, 2013, and 2015, respectively (Table~\ref{ellipses}), indicating the same outward motion of the masing cloudlets. 

For G23.389$+$00.185, 128, 118, and 129 maser spots were found in 2004, 2013, and 2015, respectively. Their elliptical fit shows an elongation towards the NE-SW direction (Fig.~\ref{threeepochs}). Seventy-nine spots appeared at all three epochs and are grouped in ten cloudlets. They all
showed linear motions and were used to derive the average proper motion relative to the brightest cloudlet at the LSR velocity of 75.338~km~s$^{-1}$. The result is presented in Fig.~\ref{g23389cm}. Proper motions range from 0.02 to 0.7~mas~yr$^{-1}$, corresponding to 0.5 to 16.1~km~s$^{-1}$ at a distance of 4.8~kpc \citep{r19}. Between method~1 and method~2, the results are consistent. The major axis of the best-fitted
ellipses increased in time by ca.~30 mas over 10.5~yrs. The radial and tangential components of proper motions calculated in method~1 are presented in Fig.~\ref{fig_rot_exp}. Again, the outward radial motion dominates, and its mean is 3.4$\pm$0.4~km~s$^{-1}$. This source shows little variation in the 6.7~GHz methanol maser emission in recent years (Fig.~\ref{spectravar}), as reported by \cite{sz18}. However, we detected an overall decrease in the flux density of the brightest spectral features, particularly between 2004 and 2013, and also of the redshifted spectral features (Fig.~\ref{threeepochs}). Moreover, the source has recently undergone an outburst, reported by \cite{t23}. The spectral features at the LSR velocities of 74.7 and 75.3~km~s$^{-1}$ have increased their flux densities to 100 and 120~Jy, respectively. Interferometric observations are needed to properly identify the outbursting region.

In our sample, G23.657$-$00.127 is the target with the most circular and widespread distribution of methanol maser spots. Details were described in \cite{b20}, and we summarise them here for completeness of the sub-sample. 
Radially outward motions were clear, with values from 0.5 to 5.4~km~s$^{-1}$ in the sky. Consistent results of radially outward motions were obtained by applying either method~1 or method~2 when we excluded the blueshifted cloudlet to the SW for the ellipse fit. The major axis of the best-fitted ellipses increased about 7~mas over 10.5~yr (Table~\ref{ellipses}). Again, the single-dish monitoring of the 6.7\,GHz methanol maser spectrum showed non-variable emission over nearly 20~yr within an accuracy of 10\%. \citep{sz18}. However, our EVN observations revealed significant changes in the intensity of the individual cloudlets over the whole range of velocity.

For G28.817$+$00.365, 36, 67, and 80 maser spots were detected in 2007, 2013, and 2015, respectively (Fig.~\ref{threeepochs}). Twenty-one of them
appeared at all three epochs and are grouped in seven cloudlets. We used five of them that show approximately linear motions
to derive the average proper motion (the excluded cloudlets are marked in Table~\ref{clouds}). The proper motions are presented in Fig.~\ref{g28817cm}.
They range from 0.07 to 0.6~mas~yr~$^{-1}$, corresponding to a range from 1.6 to 13.0~km~s$^{-1}$ for a kinematic distance of 4.6~kpc \citep{r19}. The decomposition of the proper motion velocity vectors into radial and tangential components is presented in Fig.~\ref{fig_rot_exp}. Due to the small number of cloudlets, the domination of any of these cannot be estimated. The mean radial motion is 3.5$\pm$0.4~km~s$^{-1}$. 
Gaussian profiles of maser cloudlets show that the peak of the brightest one (at the LSR velocity of 90.6~km~s$^{-1}$) decreased in time by a factor of three between 2007 and 2013. More complex spectral profiles also appeared in the LSR velocity range 91--92~km~s$^{-1}$ between 2007 and 2013. Again, this is opposite to the single-dish monitoring results (Fig.~\ref{spectravar}; \citealt{sz18}).

Towards G31.047$+$00.356, 31, 77, and 83 single maser spots were detected in 2007, 2013, and 2015, respectively (Fig.~\ref{threeepochs}).
Twenty-three spots appeared at all three epochs and are grouped in four cloudlets. Uniquely for this target, we analysed the motions of all single spots because of the significant extension of the western maser emission that would be missed when grouping spots into cloudlets. The
brightest spot was spectrally stable (at the LSR velocity of 80.709~km~s$^{-1}$) and was selected as a reference. Twenty-one spots showed linear motions and
were used for calculating the average proper motion (the two excluded spots are marked in Table~\ref{clouds}). Proper motions are presented in Fig.~\ref{g31047cm}. They range from 0.03 to 0.31~mas~yr$^{-1}$, corresponding to a range from 0.7 to 7.1~km~s$^{-1}$ for a kinematic distance of 4.9~kpc \citep{r19}. Similarly, as in G28.817$+$00.365, the decomposition of proper motion velocity vectors into radial and tangential components does not indicate a dominant expansion or rotation motions (Fig.~\ref{fig_rot_exp}). Gaussian profiles of cloudlets between 2007 and 2013 exhibit significant changes (Fig.~\ref{threeepochs}), and we did not find any non-variable feature. Also, we noticed that the cloudlet at the LSR velocity of 82.99~km~s$^{-1}$ shows the radial motion component of $-$0.10~m~s$^{-1}$~yr$^{-1}$. The single-dish monitoring from 2009 did not reveal strong variability, although there is a systematic spectral drift of the redshifted components (Fig.~\ref{spectravar}). 

Towards the source G31.581$+$00.077, 35, 43, and 49 single maser spots were detected in 2007, 2013, and 2015, respectively (Fig.~\ref{threeepochs}). Nineteen of them were identified as appearing at all three epochs. They were grouped in eight cloudlets. Six of them showed linear motions and were used to derivate average motions. Proper motions are presented in Fig.~\ref{g31581cm}, and the decomposition of them into radial and tangential components is presented in Fig.~\ref{fig_rot_exp}. We did not notice any sign of regularities in them. This source lies at 5.5~kpc, as its trigonometric parallax was measured by \cite{r19}. Again, interferometric data show amplitude variability of the spectral features, contrary to the single-dish data (Fig.~\ref{threeepochs}). 

At this point we can state that the differences in variability patterns based on VLBI and single-dish observations likely result from significant differences in angular resolution (arcmin vs. mas). Detailed studies of radial velocity drifts of single cloudlets combined with the proper motions may uniquely allow the 3D velocities to be estimated. 

\begin{table}
\centering
\caption{Parameters of flux-weighted ellipses fitted to the maser cloudlets that appeared in all three epochs. First-epoch fits are presented in Figs~\ref{g23207cm}-\ref{g31581cm}.}
\label{ellipses}
\begin{tabular}{@{}llll@{}}
\hline
\multicolumn{1}{c}{Source$^{a}$} & \multicolumn{2}{c}{Centre$^a$} & \multicolumn{1}{c}{Semi-axes; PA$_{\rm major}$$^b$}\\
\multicolumn{1}{c}{Distance} & \multicolumn{1}{c}{$\Delta$RA} & \multicolumn{1}{c}{$\Delta$Dec} &  \\
\multicolumn{1}{c}{(kpc)} & \multicolumn{1}{c}{(mas)} & \multicolumn{1}{c}{(mas)} & \multicolumn{1}{c}{(mas$\times$mas; \degr)} \\
\hline
 G23.207$-$00.377 & $-$55.817 & $+$81.207 &   128.33$\times$40.07; $-$48\\
 \multicolumn{1}{c}{4.18$^c$}& $-$56.378 & $+$82.097 & 129.48$\times$40.80; $-$48\\ 
 &$-$56.573 & $+$82.302 & 129.74$\times$40.90; $-$48 \\
\\
 \hline
 G23.389$+$00.185 & $-$39.045 & $-$74.610 & 101.84$\times$51.41, $+$52 \\
 \multicolumn{1}{c}{4.8$^c$} &$-$40.010 & $-$75.244 &  105.16$\times$51.06; $+$52 \\
 &$-$40.023 &  $-$75.514 &  105.10$\times$51.18; $+$52 \\
 \\
 \hline
 G23.657$-$00.127 & $-$66.058 & $-$96.411 & 134.89$\times$123.94; $-$20 \\
\multicolumn{1}{c}{3.19$^d$} & $-$66.379 & $-$96.246 & 136.01$\times$124.49; $-$19 \\
 & $-$66.824 & $-$97.350 & 136.06$\times$125.27; $-$19 \\
 \\
 \hline
 G28.817$+$00.365 & $-$47.443 &  $-$14.272 & 61.53$\times$26.47; $+$53 \\
 \multicolumn{1}{c}{4.6$^e$}& $-$46.509 & $-$13.680 & 59.77$\times$27.41; $+$53 \\
 & $-$46.429 & $-$13.658 & 59.65$\times$27.61; $+$52 \\
 \\
 \hline
 G31.047$+$00.356$^\star$ & $-$31.068 & $-$11.355 & 34.28$\times$12.57; $+$63 \\
 \multicolumn{1}{c}{4.9$^e$}& $-$30.665 & $-$10.780 & 33.54$\times$14.04; $+$63 \\
& $-$30.855 & $-$10.639 &  33.76$\times$14.07; $+$64 \\
\\
\hline
G31.581$+$00.077 & $+$46.437 & $+$69.870 & 108.22$\times$76.09; $+$87\\
 \multicolumn{1}{c}{5.5$^c$} & $+$46.626 & $+$69.040 & 108.18$\times$74.62; $+$86 \\
 & $+$46.756 & $+$68.912 & 108.18$\times$74.55; $+$86\\
 \hline
\end{tabular}
\tablefoot{$^a$ The relative coordinates to the brightest spot position in each target in the first epoch as given in Table~\ref{table1}. $^b$ The position angle of the major axis (north to east). 
Distances were derived either through the trigonometric parallaxes, $^c$ \cite{r19} and $^d$ \cite{b08}, or the calculated near kinematic distances, $^e$ \cite{r19}; the online calculator is available at http://bessel.vlbi-astrometry.org/node/378. $^\star$ The fitting was done to all maser spots visible at each epoch separately (Fig.~\ref{threeepochs}).}
\end{table}

\section{Discussion}

\subsection{Proper motions: Expansion}
In the six targets presented here, the overall distributions of methanol maser emission along ring-like structures have generally been stable for up to eight to ten years. We detected the variability of individual maser cloudlets, although the single-dish monitoring suggested non-variable emission overall. The proper motions of single maser cloudlets have been estimated to range from a maximum of 13.5~km~s$^{-1}$ to 0.5~km~s$^{-1}$. These values are in agreement with other studies of proper motions of the 6.7~GHz methanol masers (e.g. \citealt{s10a}). It can be clearly seen, especially in the case of the targets with a significant number of spots (Figs \ref{g23207cm}-\ref{g31581cm}), that the proper motions are consistent with expansion rather than infall, as was claimed for two well-known MYSOs:  Cep~A (\citealt{t11}; \citealt{s14}, \citealt{s17}) and AFLG~5142 \citep{g11}.

The best elliptical fits to the maser distribution in G23.207$-$00.377, G23.389$+$00.185, and G23.657$-$00.127 indicate that the sizes of their major axes have increased in time (Table~\ref{ellipses}). This result is consistent with our independent measurement of the maser proper motions (method~1 in Sect.~2.1), providing support to the idea that the maser distribution expands from a common centre and is part of a coherent kinematic structure.
We also note that the displacements of single maser spots lying outside the best-fitted ellipses are consistent with expansion as well. In the case of the remaining three targets, where a smaller number of maser spots (and cloudlets) were detected, the increase of ellipse axes is not obvious. Also, the proper motion studies using the method~1 (Sect.~2.1) do not imply clear expansions in these cases.


\begin{figure}
\includegraphics[scale=0.52]{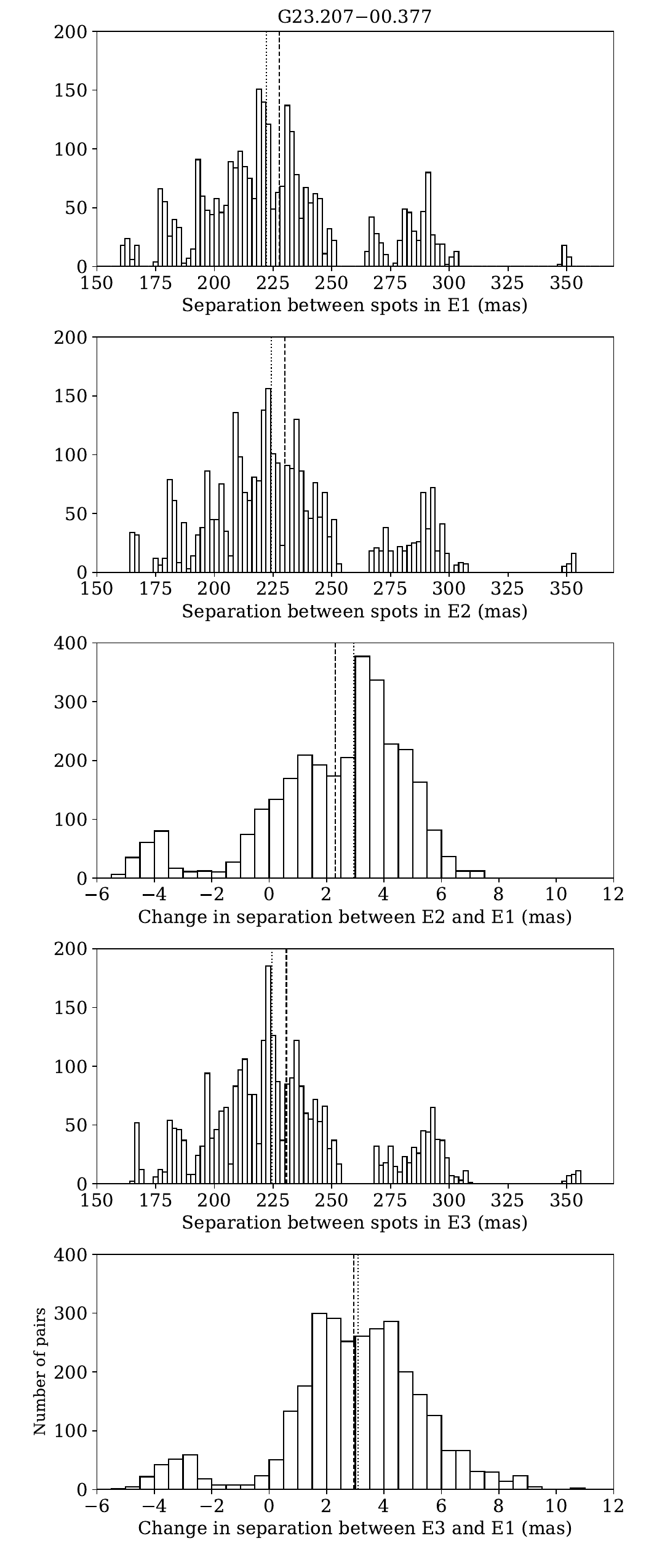}
\caption{Histograms of separations between SE and NW maser spot pairs in each epoch and the separation increase between epochs E2 and E3 relative to E1 in G23.207$-$00.377. The dashed lines mark the mean histogram values, while the dotted lines mark the median values. The histograms for the remaining targets are presented in Fig.~\ref{histogramsapp}.} 
\label{histograms}  
\end{figure}

In order to quantify the internal motion of maser spots in each source in a manner independent from the methods described above, we calculated the separations between pairs of spots from distinct regions at each epoch. In Figs~\ref{histograms} and \ref{histogramsapp}, we present histograms of separations between given groups (the separated ones) of cloudlets in each epoch. For clarity, we also present the changes in these separations, that is, the differences between epochs: i) 2004 (or 2007) and 2013, and ii) 2004 (or 2007) and 2015. The mean and median values are also marked with dashed and dotted lines, respectively. 

In the case of G23.207$-$00.377, we used 72 blueshifted maser
spots from the SE part and 40 spots from the NW  (selected at the intermediate LSR velocities). The mean and median values of the spots' separations increased 
by 2.31 and 2.95~mas, respectively, between 2004 and 2013, and by 2.95 and 3.09~mas between 2004 and 2015. An increase in the spots' separation of 3~mas corresponds to velocities of 7.18~km~s$^{-1}$ (for the time baseline 2004-2013) and 5.77~km~s$^{-1}$ (for 2004-2015) at the distance of the source (Table~\ref{ellipses}).  

In G23.389$+$00.185, we calculated the separation between pairs of spots from the NE (32 spots) and the SW parts (27 spots). The mean and median values
also increased (2.5 and 1.6~mas for separations between 2004 and 2013, and 2.9 and 2.0~mas between 2004 and 2015, respectively). For this source,
an increase in the spots' separation of 2~mas corresponds to 5.5~km~s$^{-1}$ (2004-2013) and 4.4~km~s$^{-1}$ (2004-2015). 

In G23.657$-$00.127, we considered pairs of spots lying to the north and south of the EW line at $\Delta$Dec$= 20$~mas (see Fig.~1 in \cite{b20}).
The northern group contained 88 spots, and the southern one had 120 spots. The mean and median values of their relative separation increased by 1.66 and 1.70~mas
for the time baseline 2004-2013 and by 2.13 and 2.15~mas between 2004 and 2015. At the distance of the source, an increase of 2~mas corresponds to velocities
of 3.7~km~s$^{-1}$ (2004-2013) and 2.9~km~s$^{-1}$ (2004-2015). 

In G28.817$+$00.365, we considered 19 spots located in the eastern part and two spots from the western part. The mean (median) values of changes in separations are 0.82 and 0.72~mas (0.86 and 0.63~mas) for a time baseline of six and eight years, respectively. An increase in the spots' separation of 1~mas from 2007 to 2013 or from 2007 to 2015 corresponds to velocities of 4.0~km~s$^{-1}$ and 2.9~km~s$^{-1}$, respectively. 

Similarly, in 31.047$+$00.356 we calculated the separations between pairs of spots at east and west, and their mean and median values of changes in separations are as follows: 0.64 and 1.44~mas (2007-2013) and 1.08 and 2.08~mas (2007-2015), respectively. Here, a displacement of 1~mas from 2007 to 2013 or from 2007 to 2015, corresponds to velocities of 3.8~km~s$^{-1}$ and 2.8~km~s$^{-1}$, respectively. 

Finally, in G31.581$+$00.077, we calculated the separations between spots from the east (seven) and west (nine spots) regions. The means of their changes are $-$0.11 and $-$0.03~mas, and the median values are 0.02 and $-$0.01~mas for 2007-2013 and 2007-2015, respectively. 

The separation analysis confirmed that for sources G23.207$-$00.377, G23.389$+$00.185, and G23.657$-$00.127, a clear overall expansion is detected. In addition, a tentative expansion is seen in G28.817$+$00.365 and G31.047$+$00.356. For these two latter sources, only a small number of spots was available, and the time baseline of the observations was shorter, so the results are less robust and need to be confirmed with future observations. In the case of G31.581$+$00.077, which is the most distant target from our sample, we cannot exclude a sign of rotation considering the separation analysis and the proper motion studies. The fitted ellipses have similar axes in three epochs of observations as well. Therefore, this scenario has to be confirmed by future observations and repeated proper motion studies with a longer time baseline. 

\subsection{Infrared and radio counterparts}
In general, it is difficult to relate the 6.7~GHz maser spatial morphology, which we measured at the mas scale, to the position of the young star, which can be inferred, for instance, from infrared and radio continuum observations at scales more than an order of magnitude larger so far. In the following, we collect the information available in the literature for the five sources.

Inspection of mid-infrared data from the {\it Spitzer} IRAC maps, GLIMPSE, and MIPSGAL\footnote{https://irsa.ipac.caltech.edu/} revealed that three out of the six targets (i.e. G23.207$-$00.377,
G23.389$+$00.185, and G23.657$-$00.127) coincide with unresolved sources. They are also so bright that they saturate the detectors; the coincidence is within a pixel (i.e. 1$\farcs$2) in a GLIMPSE map. \cite{deb12} presented high-resolution near- and mid-infrared images of G23.389$+$00.185 and G23.657$-$00.127. The emission
peaks are located northward (ca.~0\farcs25) from the methanol maser ring G23.389$+$00.185. Similarly, the near- and mid-infrared peaks are slightly shifted
southward from the ring G23.657$-$00.127. In both cases, there was no compelling evidence in support of the hypothesis that methanol masers resided in
circumstellar discs. \cite{h2016} also reported weak (0.38~mJy) continuum emission at 6~GHz towards G23.389$+$00.185, where the
emission shows an elongated morphology that could be interpreted as a jet associated with the methanol masers. However, the poor astrometry and sensitivity
of the radio continuum observations did not allow for a detailed comparison of position and morphology with the masers. Furthermore, \cite{b09} reported a non-detection of 8.4~GHz continuum emission towards these three sources, with an upper limit of 150~$\mu$Jy~beam$^{-1}$.

In contrast, the 6.7~GHz methanol masers in G28.817$+$00.365 were found within 0\farcs1 of the position of the 8.4~GHz continuum peak \citep{b09}. This emission, likely an ultra-compact H{\small II} region, was also imaged at 6~GHz by \cite{h2016}.
A search for mid-infrared data from the catalogues mentioned above did not show any clear counterpart. Similarly, the G31.047$+$00.356 masers are offset
from the mid-infrared sources. \cite{h2016} reported weak continuum emission at 6~GHz that likely coincides with the methanol emission; this
emission was previously undetected at 8.4~GHz by \cite{b09} with an upper limit of 150~$\mu$Jy~beam$^{-1}$. In the case of G31.581$+$00.077, a continuum source was detected by \cite{h2016} but with no evidence supporting the physical relationship between the maser and continuum source since the maser is offset eastward by 0.19~pc. 

We searched the Atacama Large Millimeter/Submillimeter Array (ALMA) Archive and found three sources with available data: G23.389$+$00.185, G23.657$-$00.127, and G31.581$+$00.077. They were part of the large programme ALMAGal (ALMA Evolutionary study of High Mass Protocluster Formation in the Galaxy; ID: 2019.1.00195.L). The observations covered the frequency range from 216.8\,GHz to 244.14\,GHz (band 6), with the best spatial resolution being 0.27" for G23.389$+$00.185 and G23.657$-$00.127 and 5.57" for G31.581$+$00.077. Methanol masers in G23.389$+$00.185 overlap with the centres of continuum emission of C$^{18}$O and CH${_3}$CN emission, proving that the
local gas is warm and that the methanol masers arise from a hot molecular core (due to the nearby presence of CH${_3}$CN emission) and a larger gas condensation (traced by C$^{18}$O). In G23.657$-$00.127, however, we observed a westward offset by 0.006\,pc, and in G31.581$+$00.077, we observed a south-west offset by ca.~0.21\,pc (Fig.~\ref{almacounter}). A detailed analysis of the association between thermal and maser lines at high angular resolution goes beyond the scope of this study, and it is postponed to a subsequent dedicated paper. Moreover, the angular resolution of the data is diverse. It is especially poor for G31.581$+$00.077, and the offset might be marginal.

\subsection{Origin of the methanol rings}
\cite{s19} targeted the high-mass star-forming G23.01$-$0.41 with the ALMA at a resolution of 0\farcs2 and directly imaged a molecular disc around a MYSO in thermal methanol (and methyl cyanide). The disc rotates and undergoes infall around the central star and drives a molecular jet arising from the inner disc regions ($<$\,800\,AU). The environment surrounding the young star is
also one of the richest (Galactic) sites of methanol maser emission, where the maser cloudlets are distributed within a radius of 1500\,AU from the central
star and, on average, show expanding motions on the order of 7~km~s$^{-1}$ (\citealt{s15}, their Fig.\,4). In five out of the six sources presented in this paper (i.e. all but G31.581$+$00.077), the methanol maser kinematics show similar properties, with the maser emission distributed over large areas (300-800~AU) and
the proper motions tracing expansion at velocities of several kilometers per second. These findings might be consistent with a scenario where the methanol masers are excited in the inner outflow cavity at the interface with a flare disc.
A similar scenario explains the proper motions of SiO masers in Orion Source~I, the closest MYSO, where the masers arise from a wide-angle bipolar
wind emanating from a rotating edge-on disc \citep{m10}. 

In order to test whether this scenario also applies to the methanol maser rings, complementary observations of thermal tracers at high angular resolution ($\sim$0\farcs1)
are needed, which can be used to interpret the environment of methanol masers. In this respect, we have started a sensitive ($\sim\mu$Jy) search of radio 
continuum emission towards the methanol maser rings with The Karl G.~Jansky Very Large Array. These observations will be able to detect the ionised gas emission, including from radio thermal jets or hyper-compact H~{\small II} regions, excited by any B-type young star associated with the masers and will allow us to pinpoint 
the driving source of the masers at a resolution comparable with the extent of the maser rings. Moreover, we also expect to explore the origin of maser rings in combination with the accretion disc or outflow tracers based on locations of the emission peak of dust and thermal lines, such as CH$_3$CN, SiO, and C$^{18}$O, relative to the location of the maser.

For comparison, in G23.657$-$00.127 we proposed two scenarios: a maser ring related to an outflow or a wide-angle wind at the base of a protostellar jet tracing a combination of rotation around and expansion along the jet \citep{b20}. Similar decomposition of proper motion vectors for the remaining targets does not show a clearer scenario. In G23.207$-$00.377 and G23.389$+$00.185, the proper motion characteristics are similar to the ring, and one may assume the same physically connected structure. However, the rotational components are not convincing in that sense. 

\subsection{Modelling a plane of gas motion}
\cite{s17} showed that the 3D proper motion vectors of the methanol maser cloudlets in the HMSFR Cepheus~A HW2 move along a plane. The 3D velocity field of methanol maser cloudlets was known from the proper motions and LSR velocities with subtraction of the region's systematic velocity. They showed that the masers were detected within a radius ranging from 300 to 900~AU from HW2 tracing infall and rotational components. Due to the lack of complementary data for our sample of methanol maser rings, we were not able to infer the YSO position and the associated disc nor the jet orientation in order to give a more robust interpretation similar to what was presented in Cerpheus~A HW2. However, we used the same planar hypothesis to search for a similar behaviour in our sample. Using the results from method~1 (Sect.~2.1), we searched for the best plane containing the 3D velocity vectors by minimising the average ratio V$_z$/V$_t$, that is, the perpendicular (V$_z$) to the tangential components (V$_t$) of the vectors to a given plane in the space (for details, see Sect.~4.1 and Fig.~3 in \cite{s17}). The plane is identified with two angles: i$_{\rm model}$, the angle between the perpendicular to the plane and the line of sight, and PA$_{\rm model}$, the position angle (north to east) of the sky-projection of the perpendicular to the plane. In Table~\ref{table_model}, we list the best fits for our targets. 

We also calculated the inclination angles derived from the best ellipse fits (assuming the inclined disc-like structure). The inclination, i, is based on the ratio of minor to major axes in the same convention as i$_{\rm model}$, that is, the angle between the perpendicular to the plane of the best ellipse fit and the line of sight, and position angle of the minor axis, PA$_{\rm minor}$, is based on the ellipses's fits as given in Table~\ref{ellipses} (PA$_{\rm minor}$$\perp$PA$_{\rm major}$.) We note the agreement (within 20-40\degr) between the inclinations of the best plane from the above model and the plane of the ellipse for G23.207$-$00.377, G28.817$+$00.365, G31.047$+$00.356, and G31.581$+$00.077. However, in general, as a result of this comparison, that is, the model of the 3D proper motion vectors moving along a plane and the maser cloudlets tracing an inclined ring, we did not find an agreement. This may indicate that the masers are located in a ring and their motion is substantially out of the plane of the ring. For G23.389$+$00.185 and G23.657$-$00.127, the fits are mutually exclusive. In the case of the most circular configuration observed in G23.657$-$00.127, the plane shows an almost edge-on inclination, while the distribution of maser spots suggest the existence of a ring-like morphology seen close to face-on. That may indicate the weakness of a model that does not fit the positions but only the 3D velocities. However, extensive searches for G23.657$-$00.127 where several parameters of the model were changed suggest that there is a significant (dominant) contribution of the line of sight velocities to the 3D velocities or that the masers move at a large angle with the plane of the sky. That is generally consistent with a disc-wind scenario \citep{b20}, where although the flow starts from the disc, the velocities have a significant component along the disc (jet) axis.

\begin{table}
\centering
\caption{Parameters of best fits of the preferential planes of gas motion.}
\label{table_model}
\begin{tabular}{cccccc}
\hline
Source & i$_{\rm model}$ & PA$_{\rm model}$ &V$_{\rm sys}$ & i & PA$_{\rm minor}$\\
 & (\degr) & (\degr) & (km~s$^{-1}$) & (\degr) & (\degr) \\
\hline
 G23.207$-$00.377 & 51 & 165 & 77.5$^a$ & 72 & $+$42\\
 G23.389$+$00.185 & 6  &  15 & 75.5$^b$ & 60 & $-$38\\
 G23.657$-$00.127 & 81 & 285 & 80.5$^b$ & 23 & $+$70\\
 G28.817$+$00.365 & 76 & 285 & 86.9$^a$ & 64 & $-$37\\
 G31.047$+$00.356 & 101 & 200& 77.6$^a$ & 76 & $-$27\\
 G31.581$+$00.077 & 81 & 245 & 96.0$^a$ & 45 & $-$3\\
 \hline
\end{tabular}
\tablefoot{Systemic velocities are estimated using NH$_3$ emission from $^a$ \cite{w12}, $^b$ \cite{u11}.}
\end{table}

\section{Conclusions}
We report the proper motions of 6.7~GHz methanol maser sources that were earlier classified as having a ring-like morphology \citep{b09}. Three-epoch observations using the EVN and spanning over eight to ten years enabled us to detect velocities on the order of a few kilometers per second at distances from 3.19 to 5.5~kpc. We find that in G23.207$-$00.377, G23.389$+$00.185, G23.657$-$00.127, G28.817$+$00.365, and G31.047$+$00.356, the internal motions of maser cloudlets clearly suggest expansion from a common centre. In G31.581$+$00.077, the results are still marginal, and possible rotation cannot be excluded. The overall morphology of the maser emission has remained stable, although the intensities of individual maser cloudlets varied from epoch to epoch, suggesting internal instabilities. Our studies also show that the maser spot distribution alone does not allow for a direct interpretation of the ongoing scenario (e.g. if the rings trace the discs) and that multi-frequency studies at a complementary angular resolution are needed. 

\begin{acknowledgements}
AB, MS, PW, AK, MD acknowledge support from the National Science Centre, Poland through grant 2021/43/B/ST9/02008. The research leading to these results has received funding from the European Commission Seventh Framework Programme (FP/2007-2013) under grant agreement No. 283393 (RadioNet3).
\end{acknowledgements}

\bibliography{librarian}
\bibliographystyle{aa}

\begin{appendix}
\section{Additional tables and figures}

\begin{figure*}
\centering
\includegraphics[scale=0.57, trim={0 3cm 0 2cm},clip]{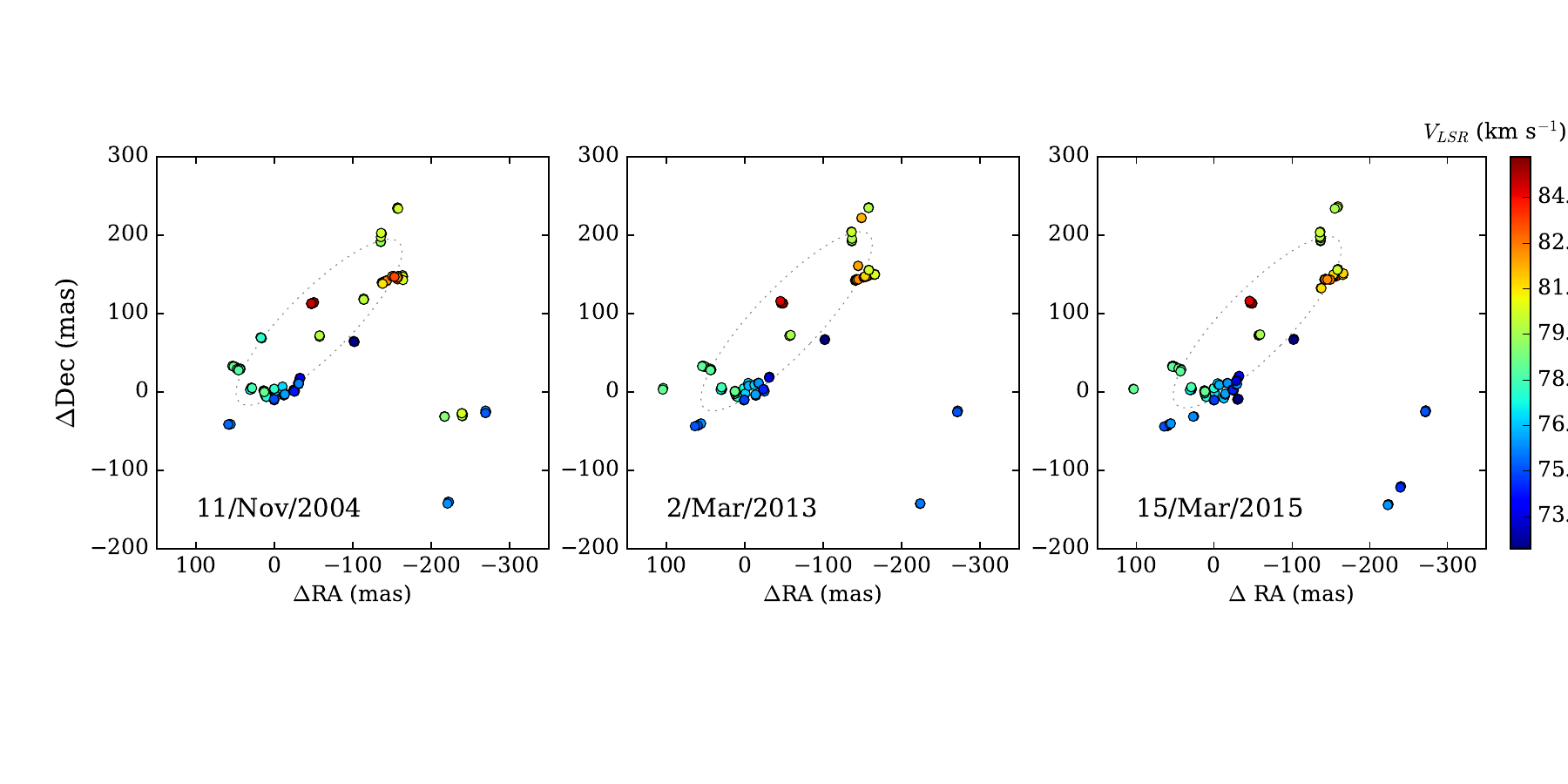}
\includegraphics[scale=0.6]{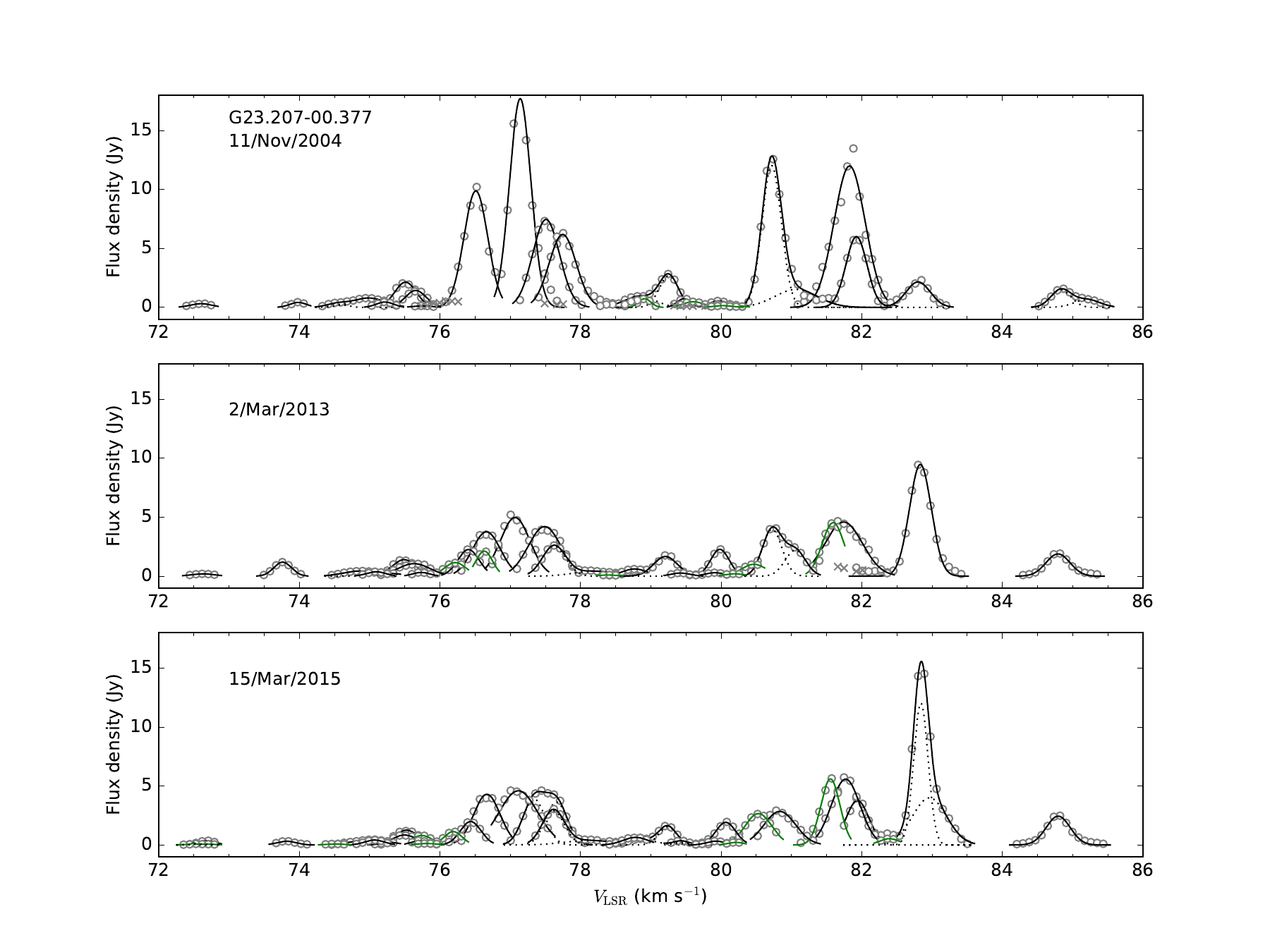}
\caption{Methanol maser emission at 6.7~GHz detected using the EVN towards G23.207$-$00.377. Top: Distribution of maser spots in each epoch. The name corresponds to the Galactic coordinates of the brightest spots listed in Table~\ref{table1}, the (0,0) locations. The colours of the circles relate to the LSR velocities as shown on the right bar. The dotted ellipses trace the best fits to all spots detected at each epoch (Table~\ref{ellipses}). Bottom: Spectra of individual 6.7~GHz maser cloudlets with Gaussian velocity profiles. Each circle traces the emission of a single maser spot, while the  lines represent the fitted Gaussian profiles. In cases with a complex profile, we also draw the single Gaussian profiles, denoted by the dotted lines. The cloudlets used for the proper motion studies are marked by the black fitting lines, and they are characterised in Table~\ref{clouds}. The green lines trace the remaining cloudlets with their Gaussian profiles, while crosses correspond to spots from cloudlets without Gaussian characteristics.}
\addtocounter{figure}{-1}
\label{threeepochs}
\end{figure*}

\begin{figure*}
\centering
\includegraphics[scale=0.57, trim={0 3cm 0 2cm},clip]{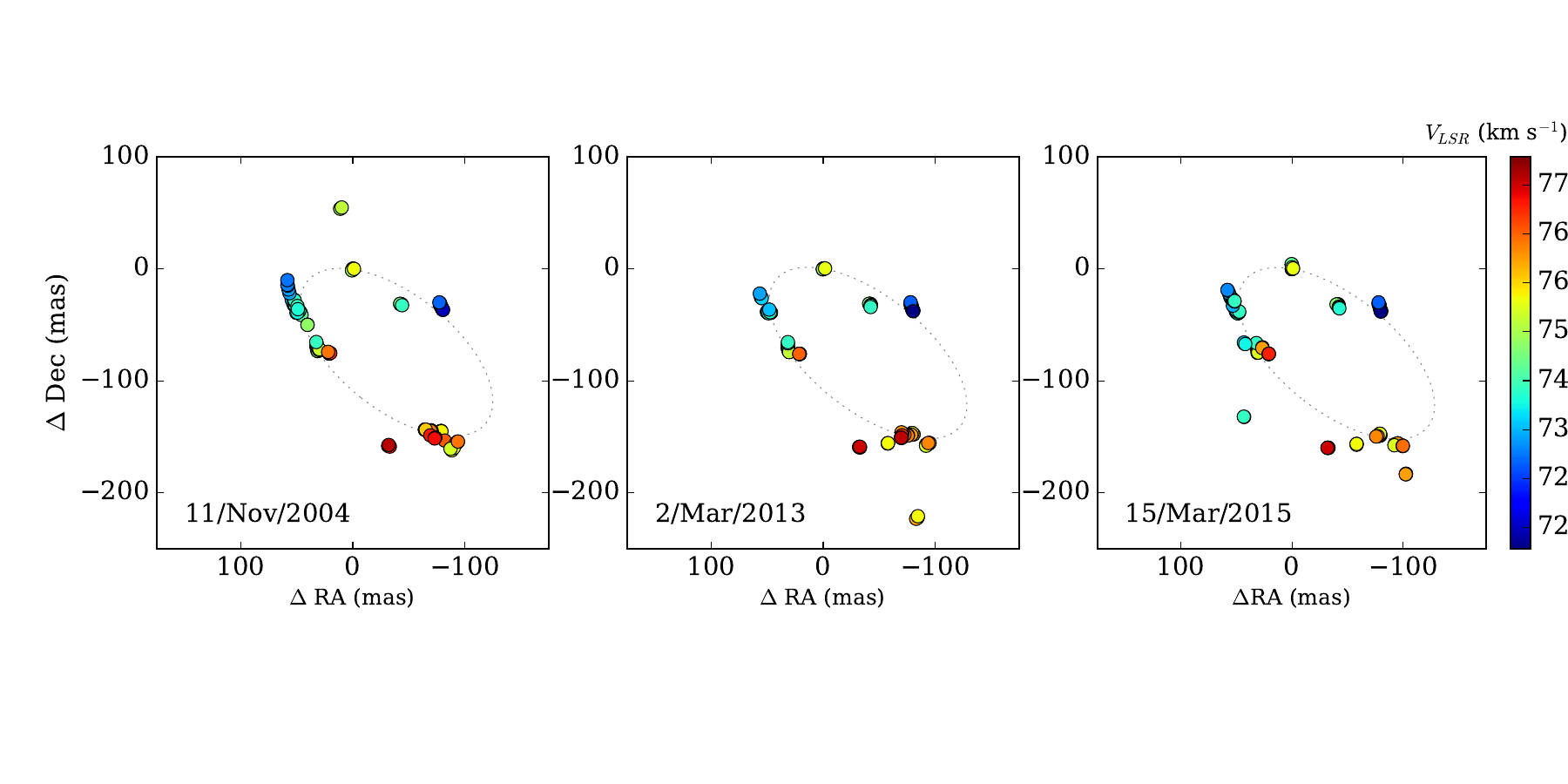}
\includegraphics[scale=0.6]{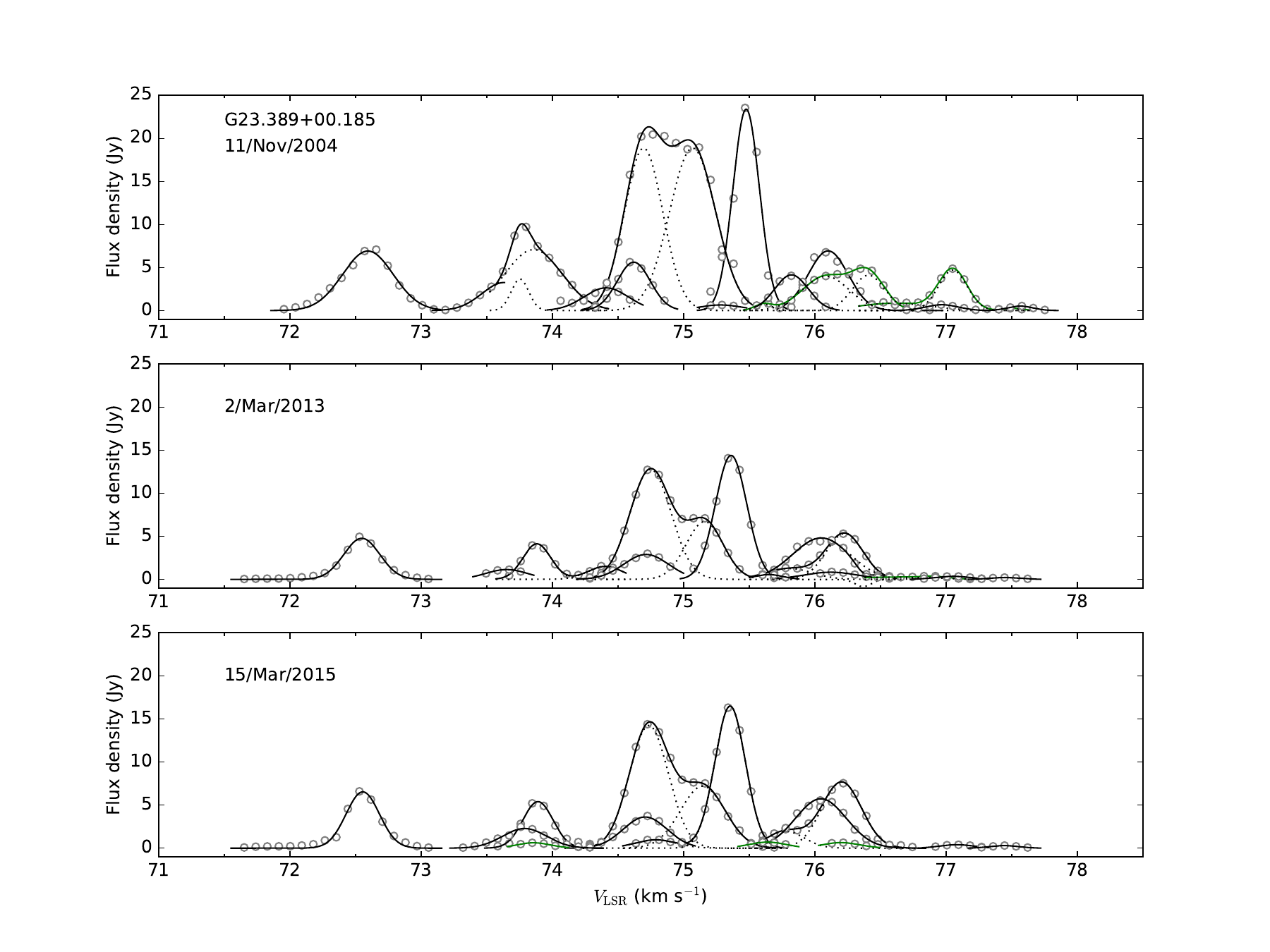}
\caption{Continued but for the G23.389$+$00.185 target.}
\addtocounter{figure}{-1}
\end{figure*}

\begin{figure*}
\centering
\includegraphics[scale=0.57, trim={0 3cm 0 2cm},clip]{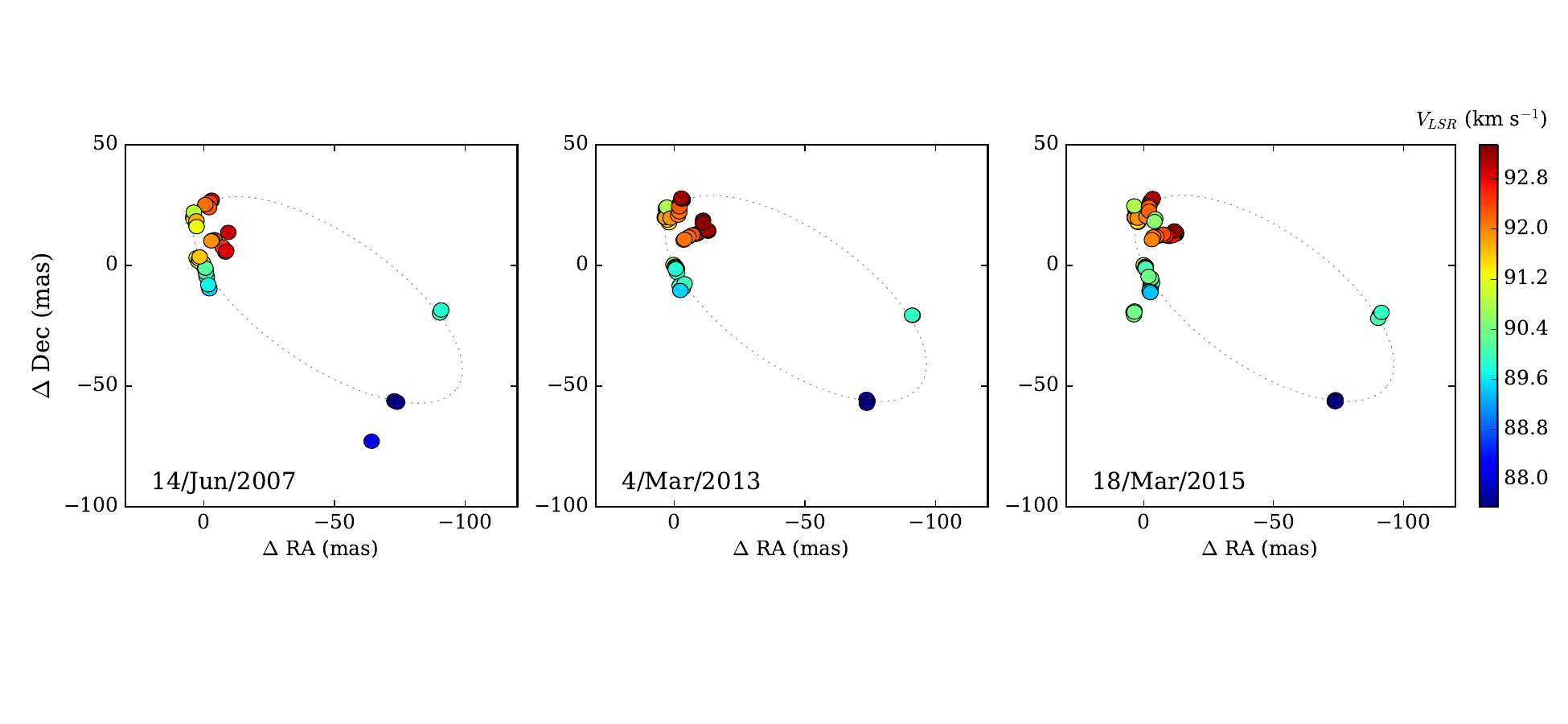}
\includegraphics[scale=0.6]{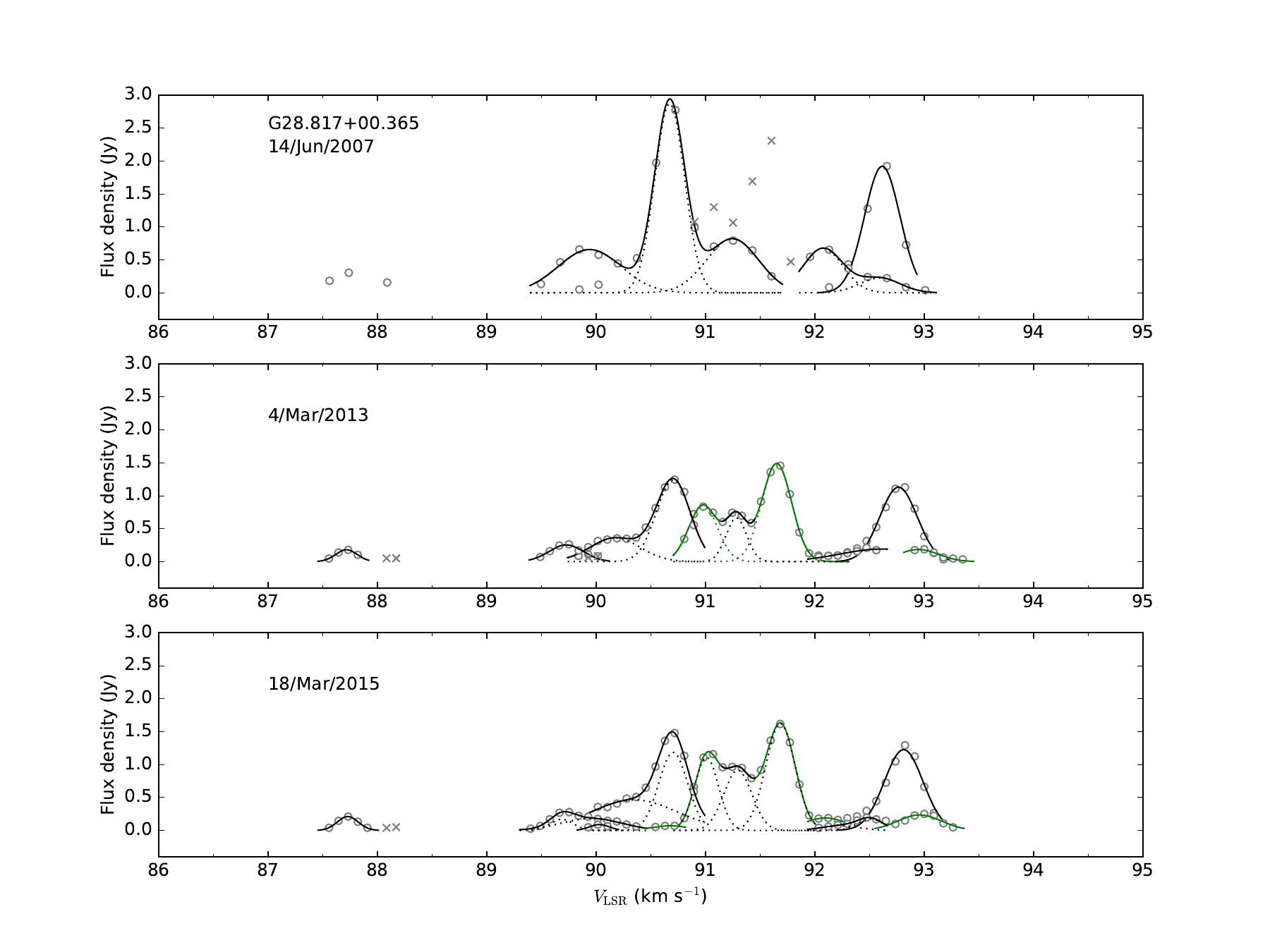}
\caption{Continued but for the G28.817$+$00.365 target.}
\addtocounter{figure}{-1}
\end{figure*}

\begin{figure*}
\centering
\includegraphics[scale=0.57, trim={0 3cm 0 2cm},clip]{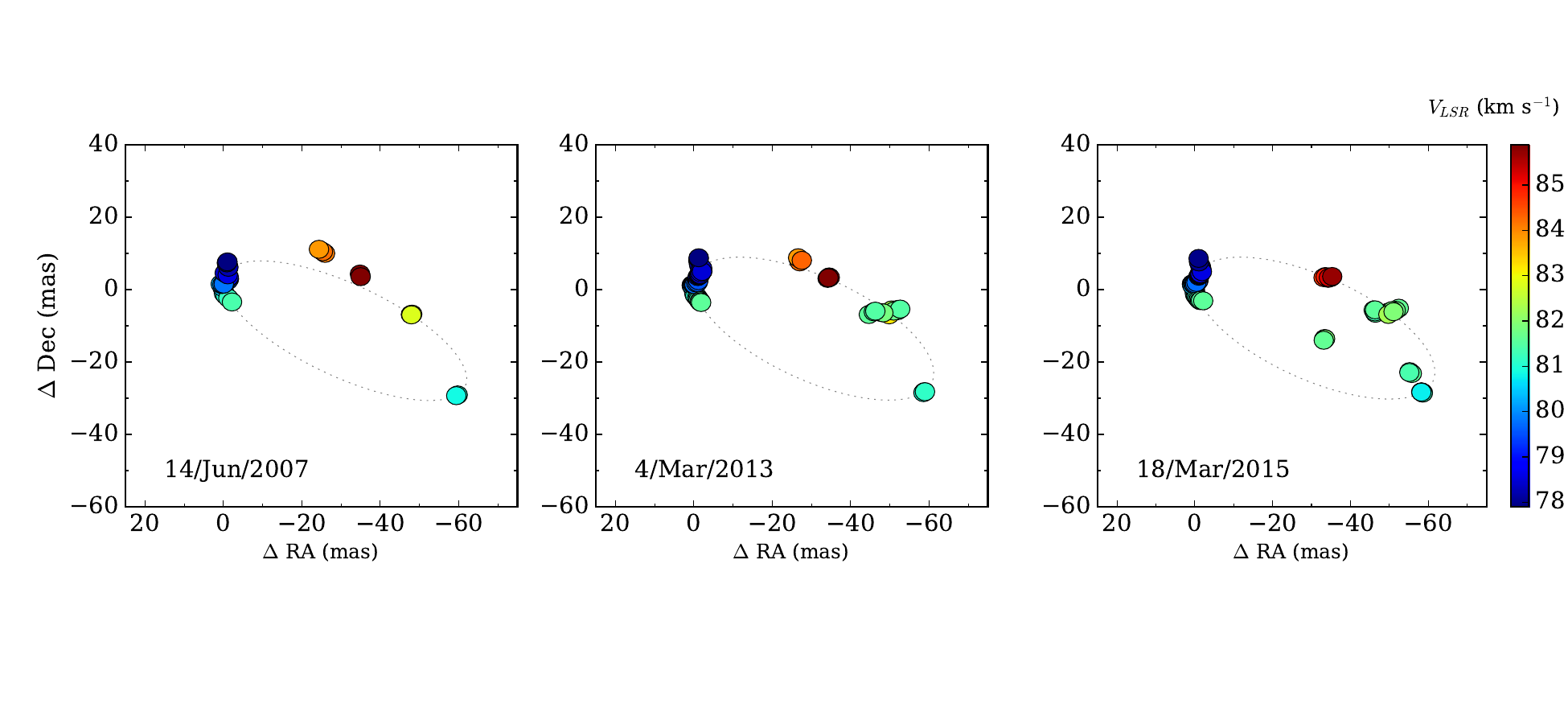}
\includegraphics[scale=0.6]{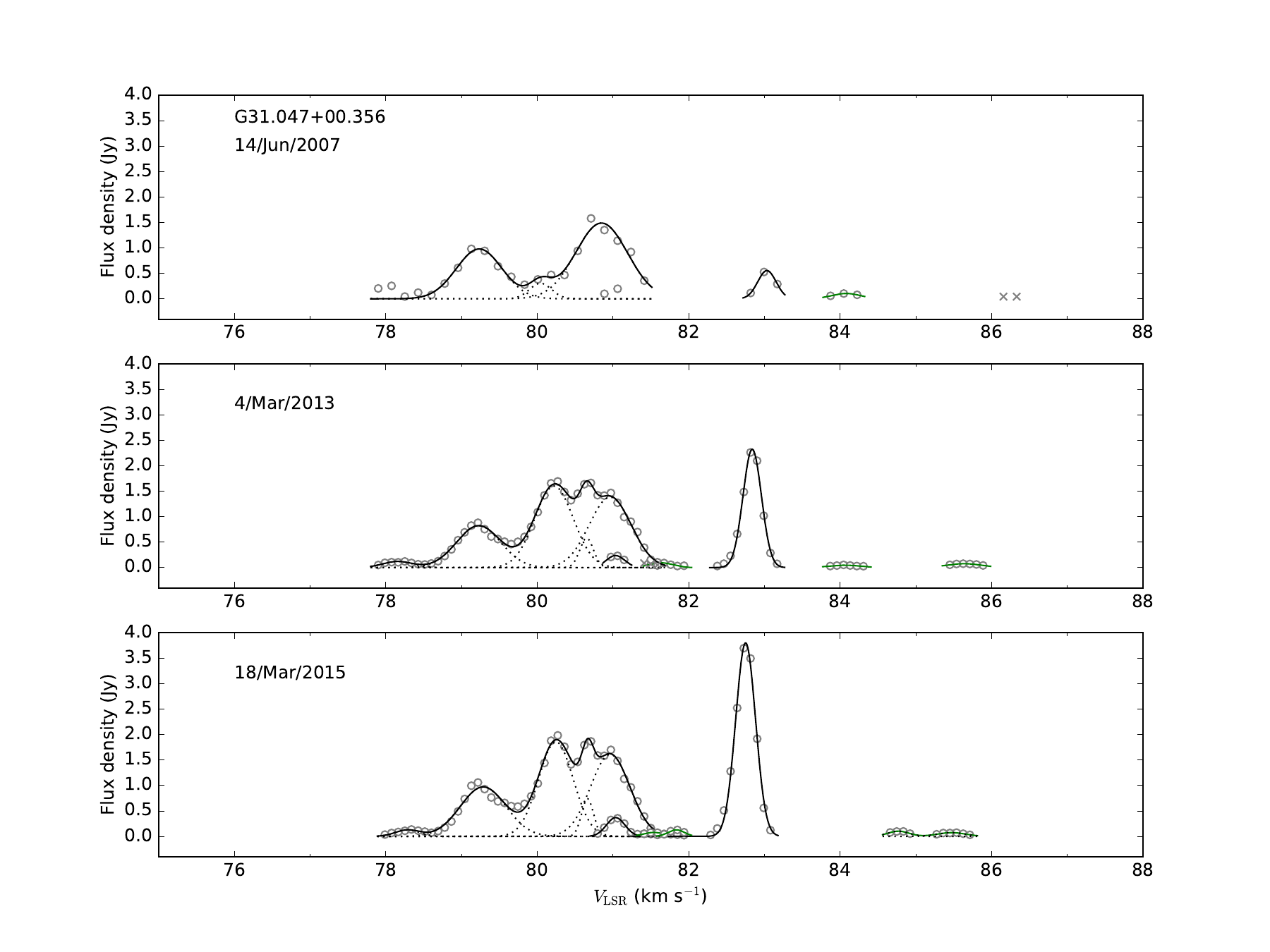}
\caption{Continued but for the G31.047$+$00.356 target.}
\addtocounter{figure}{-1}
\end{figure*}

\begin{figure*}
\centering
\includegraphics[scale=0.57, trim={0 3cm 0 2cm},clip]{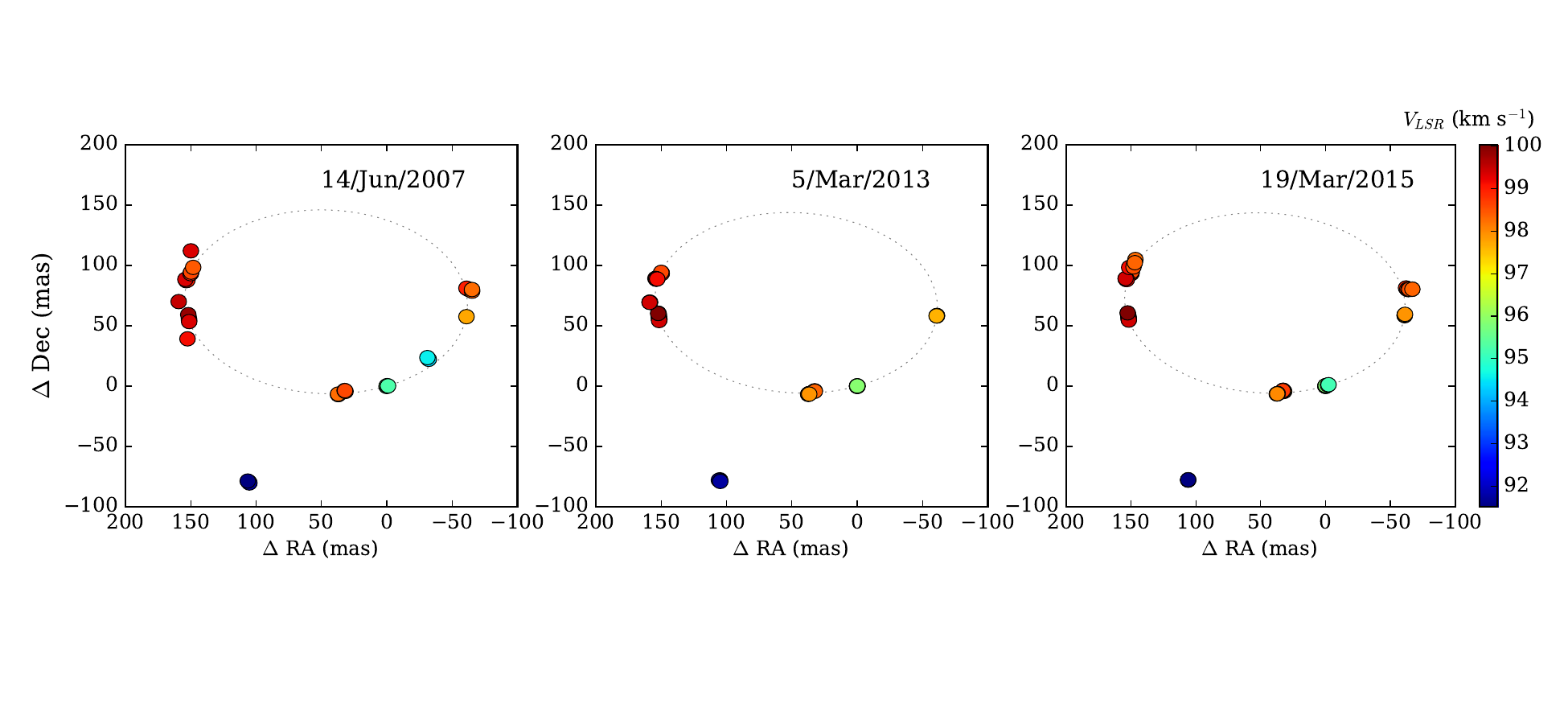}
\includegraphics[scale=0.6]{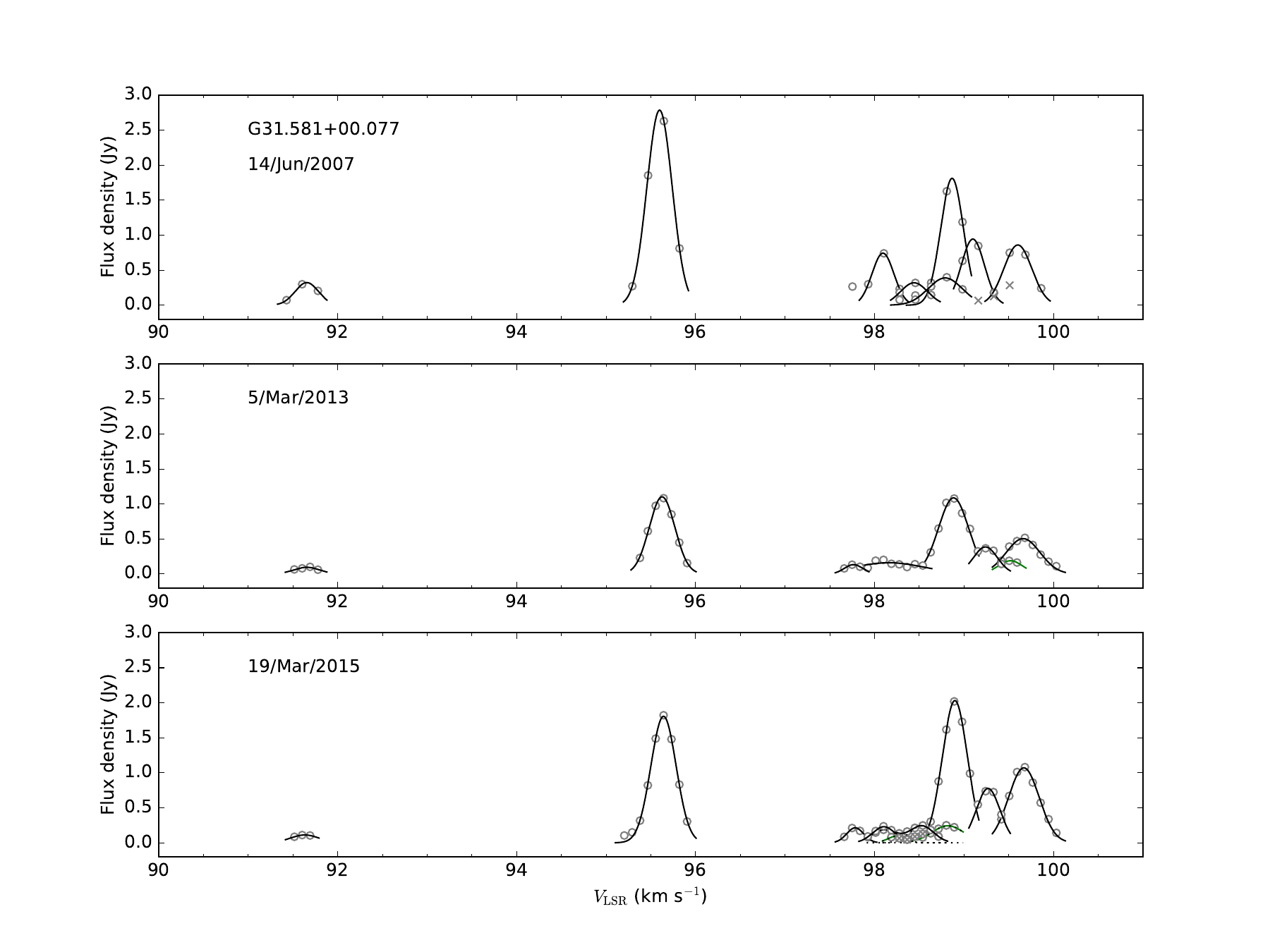}
\caption{Continued but for the G31.581$+$00.077 target.}
\end{figure*}

\begin{figure*}
\centering
\includegraphics[width=0.49\textwidth]{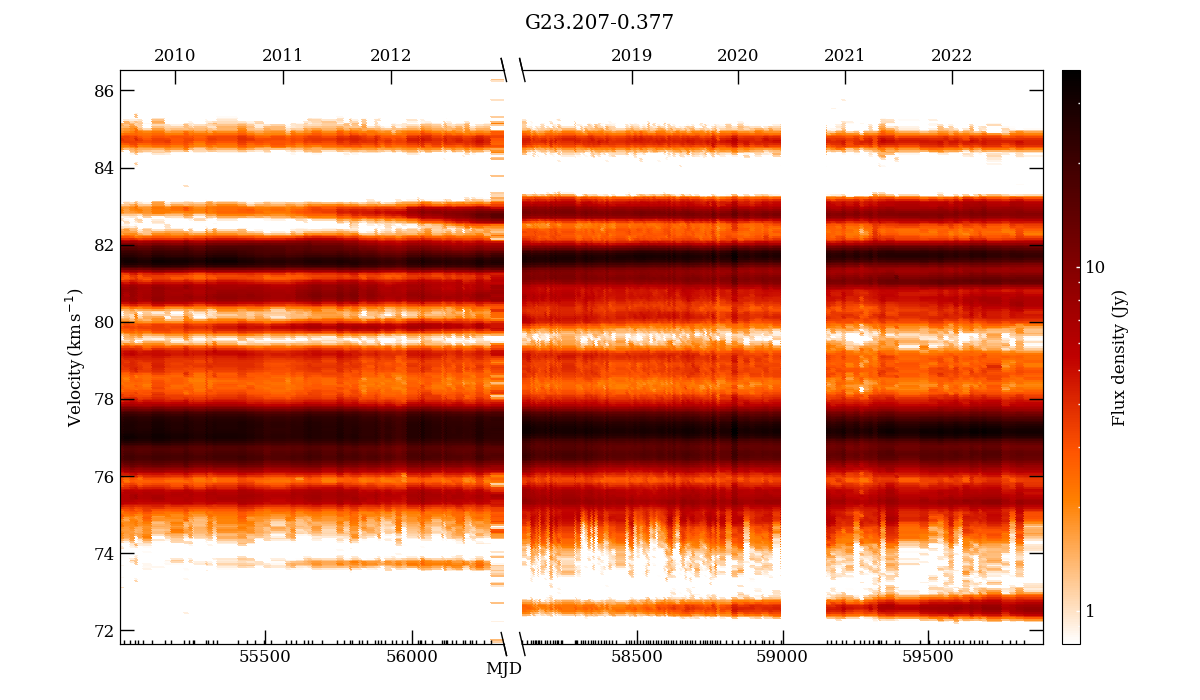}
\includegraphics[width=0.49\textwidth]{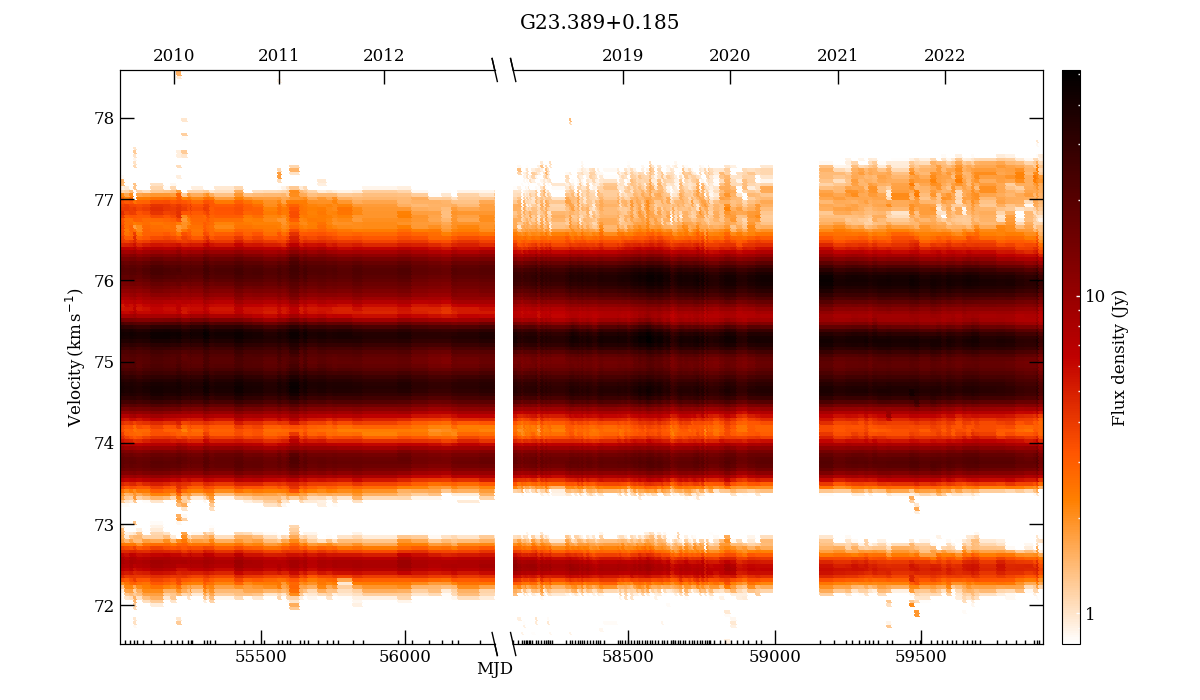}
\includegraphics[width=0.49\textwidth]{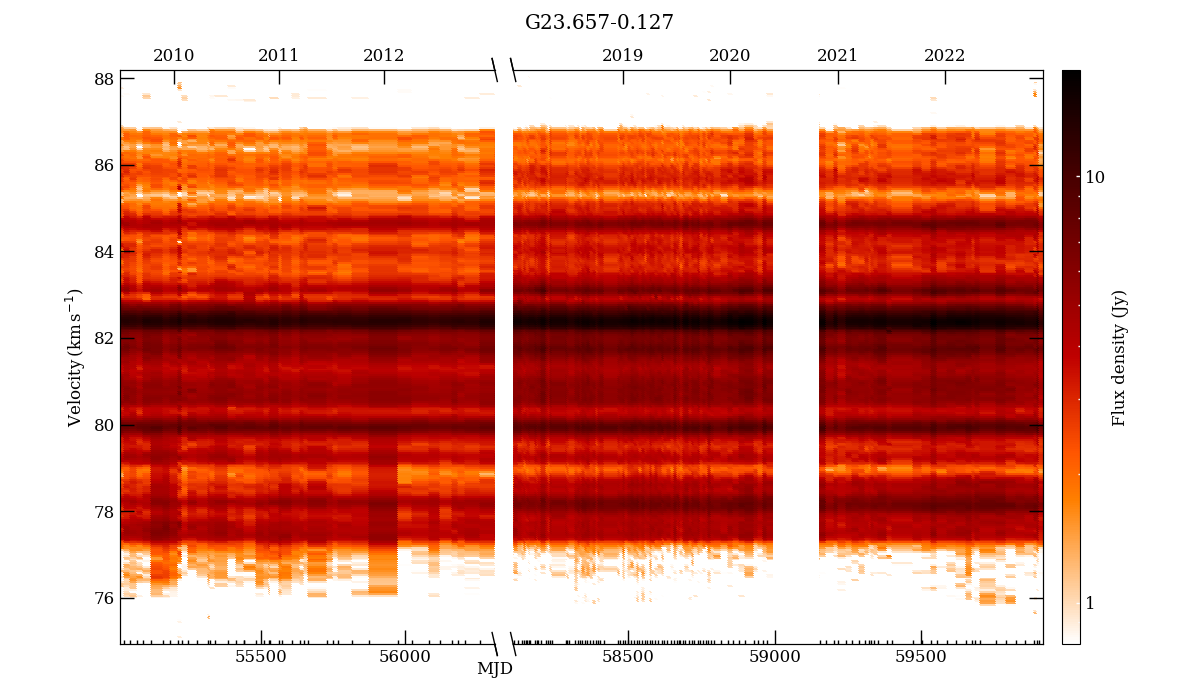}
\includegraphics[width=0.49\textwidth]{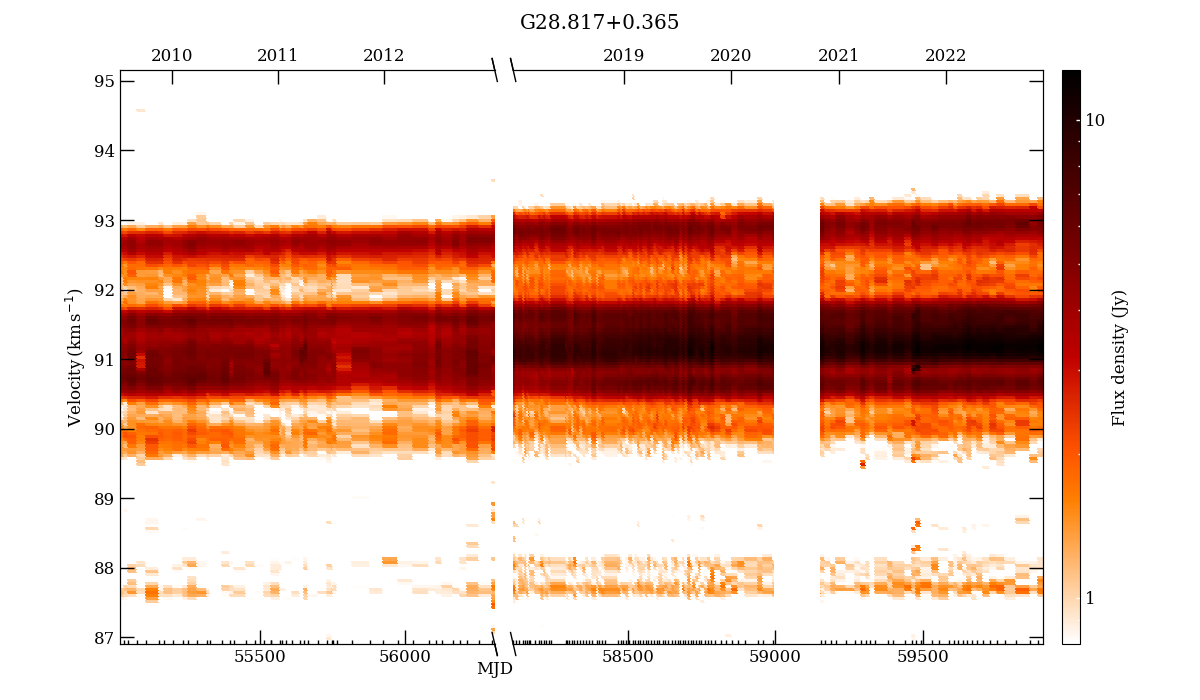}
\includegraphics[width=0.49\textwidth]{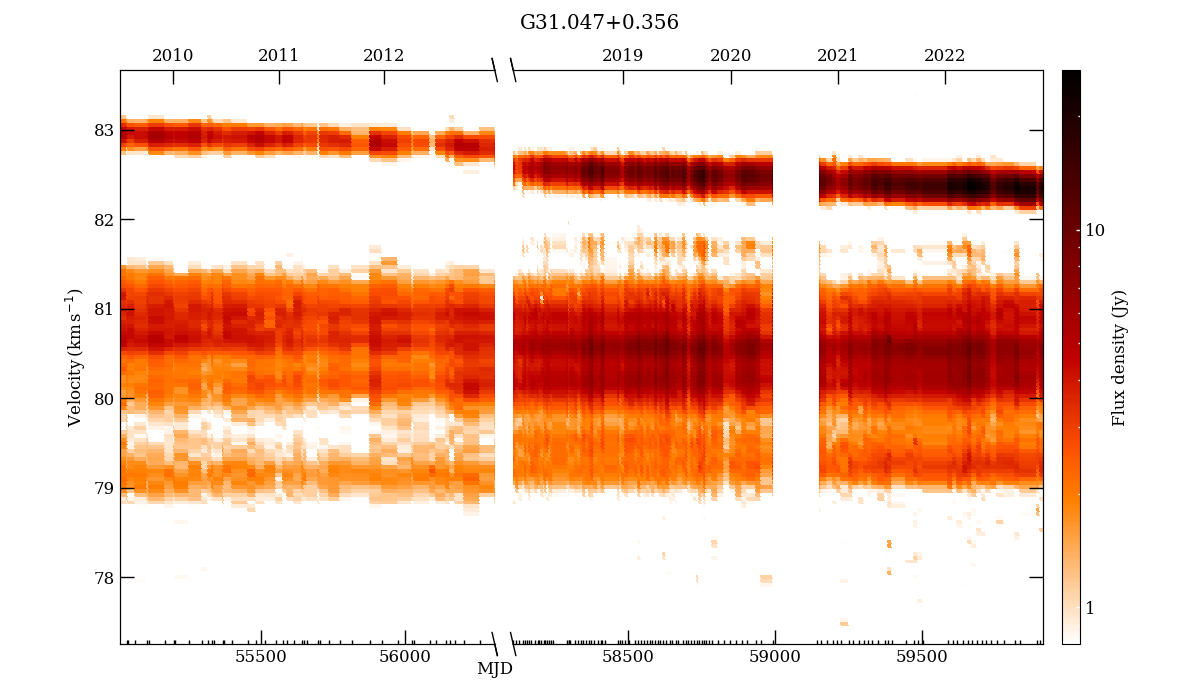}
\includegraphics[width=0.49\textwidth]{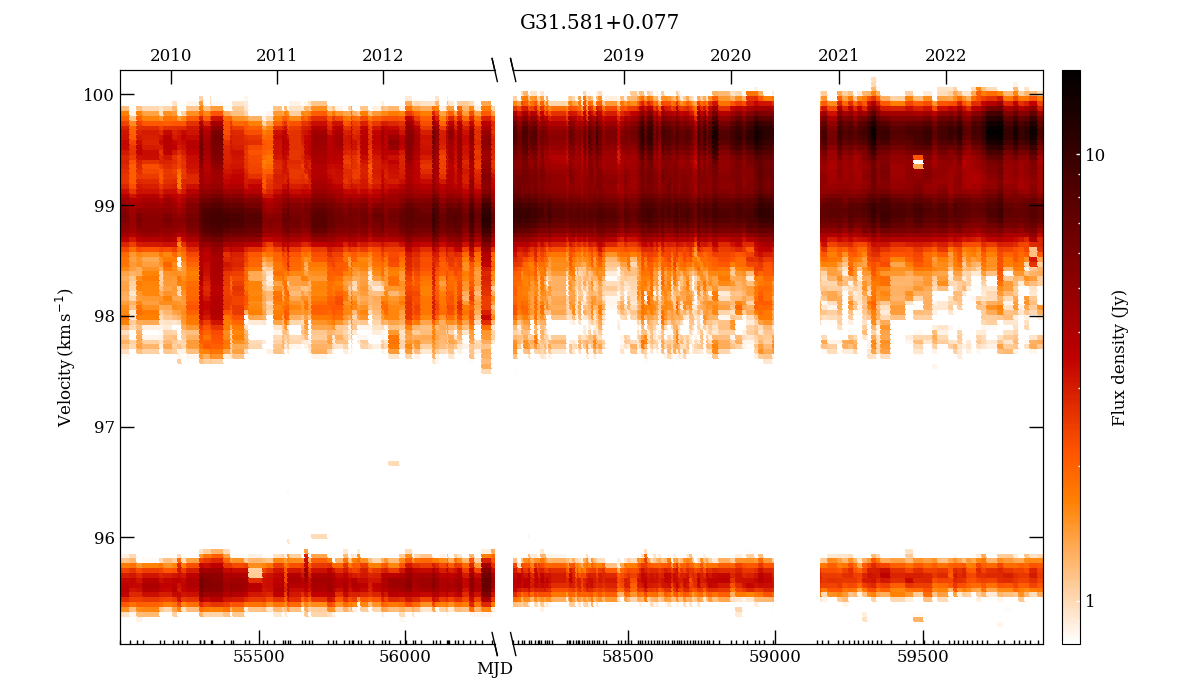}
\caption{Dynamic spectra of the variability of the 6.7~GHz methanol maser lines towards the targets as monitored using the 32-m Torun dish.}
\label{spectravar}
\end{figure*}

\begin{figure*}
\centering
\includegraphics[width=\textwidth]{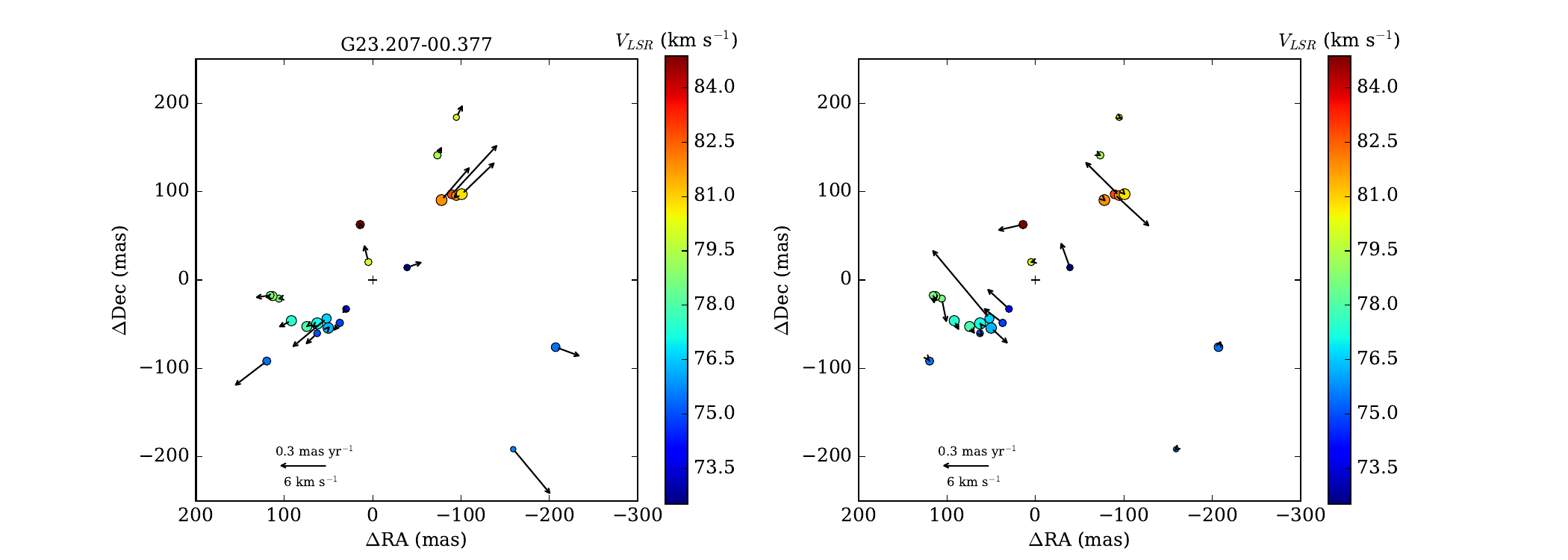}
\includegraphics[width=\textwidth]{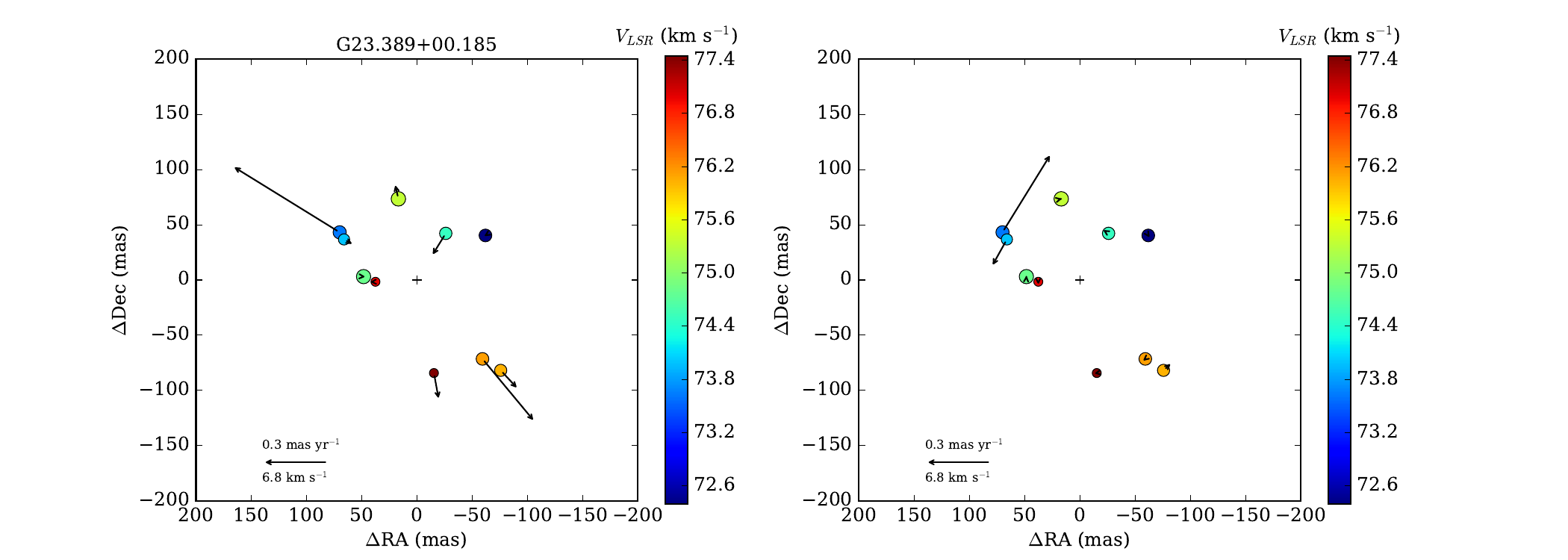}
\includegraphics[width=\textwidth]{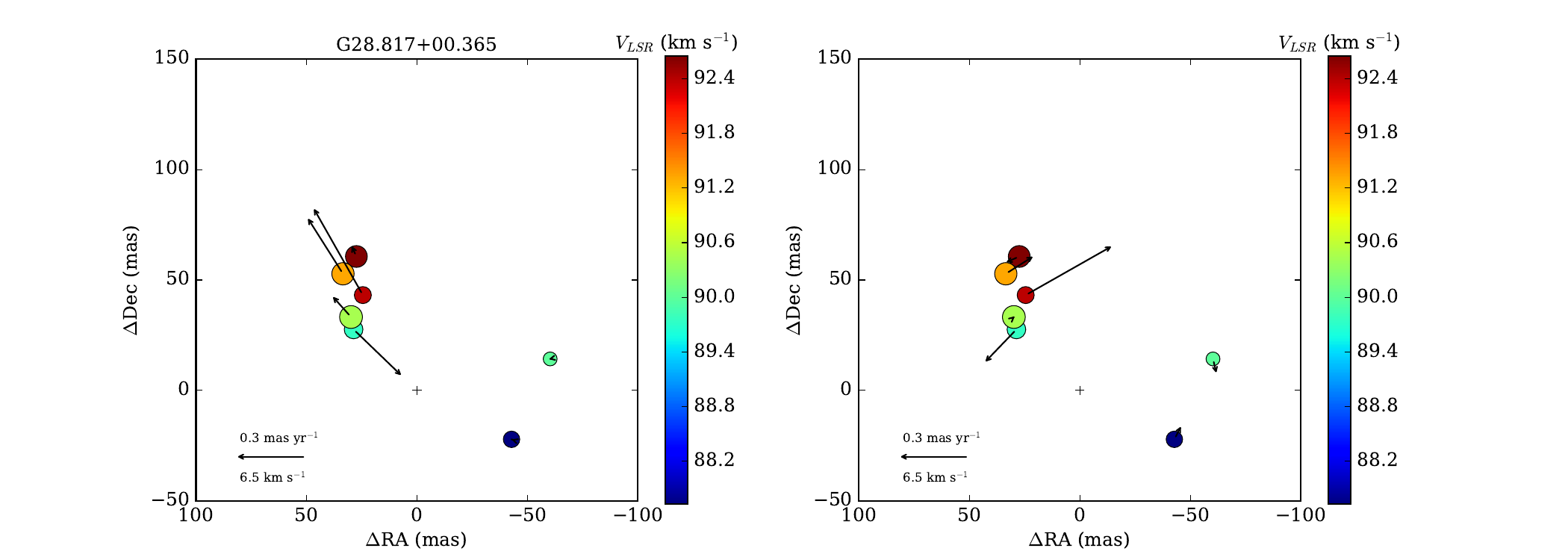}
\caption{Radial ({\it left}) and tangential ({\it right}) components of the proper motions of 6.7~GHz methanol cloudlets in targets (as labelled) calculated relative to the centre of motions marked by the plus signs.}
\label{fig_rot_exp}
\addtocounter{figure}{-1}
\end{figure*}

\begin{figure*}
\centering
\includegraphics[width=\textwidth]{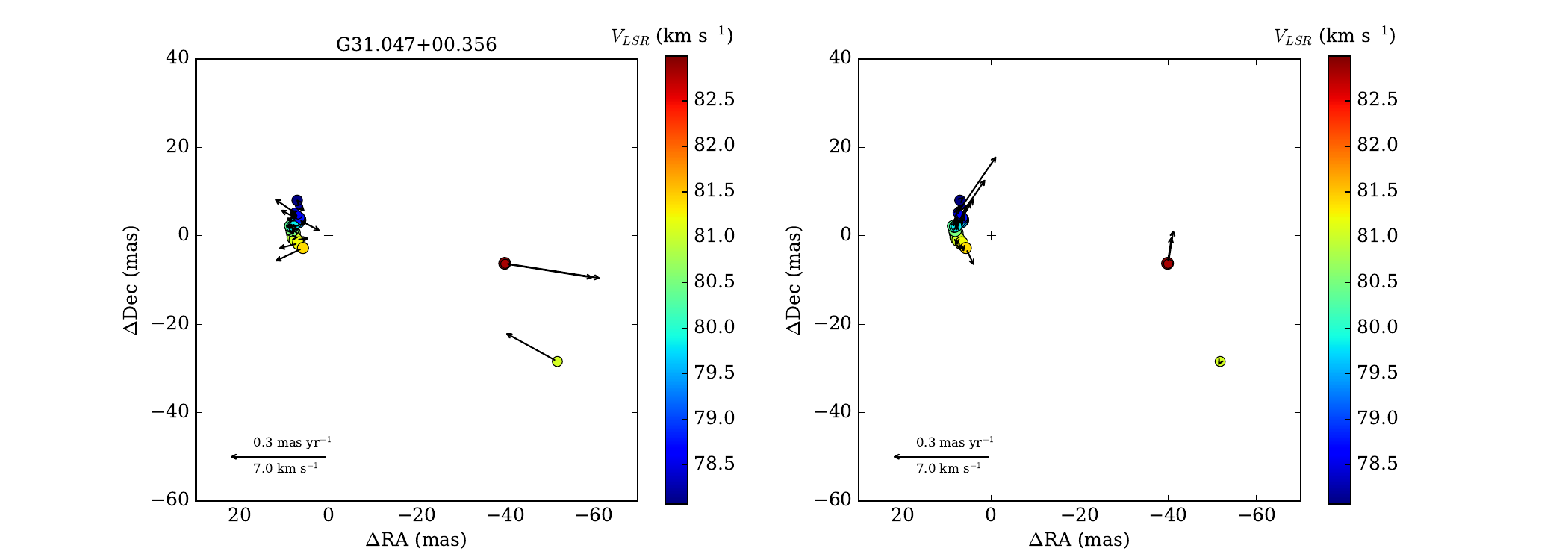}
\includegraphics[width=\textwidth]{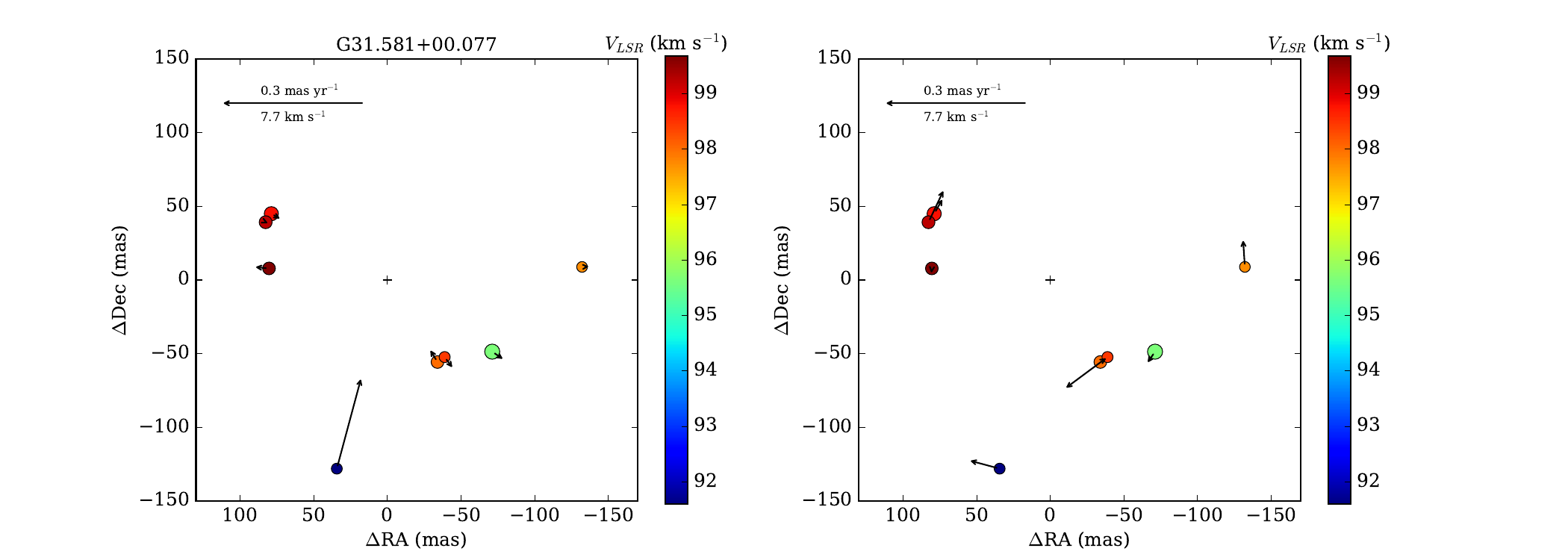}
\caption{Continued.}
\end{figure*}

\begin{figure}
\includegraphics[scale=0.52]{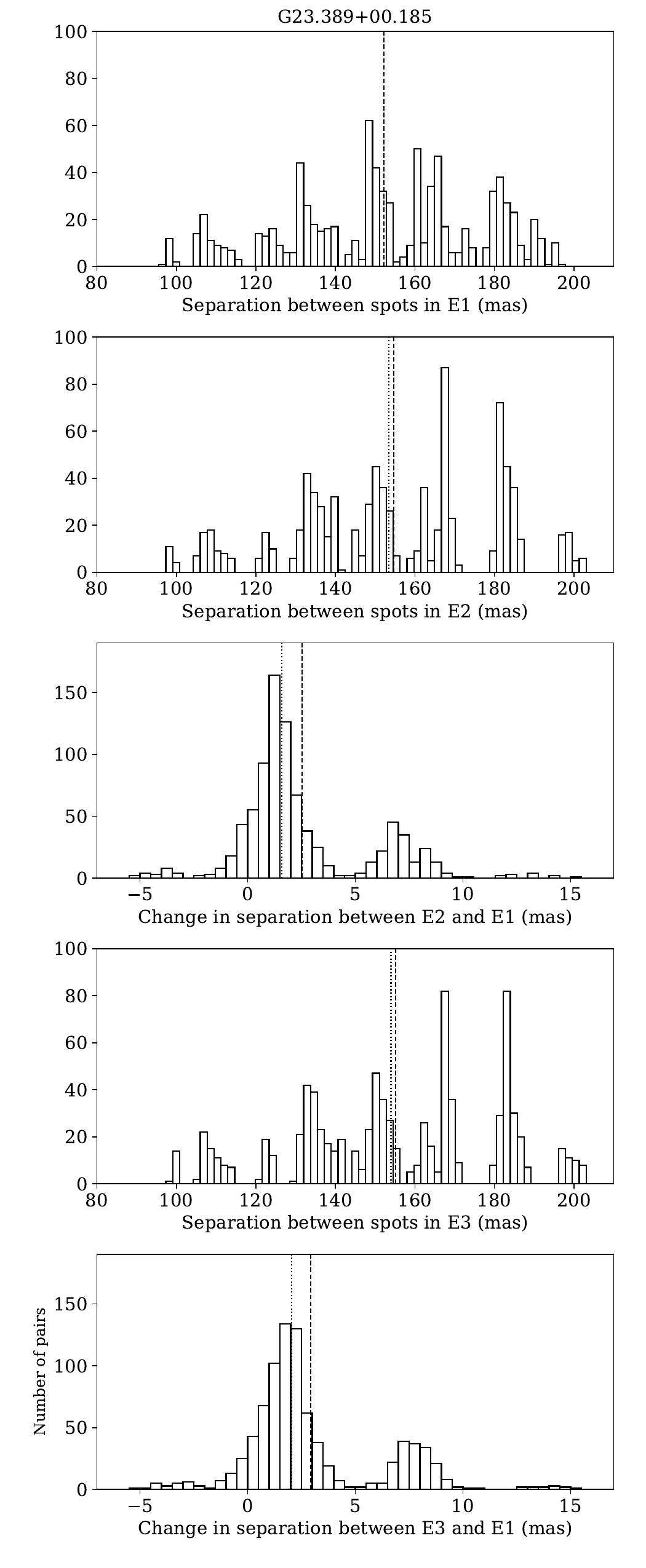}
\caption{Histograms of separations between NE and SW maser spot pairs in each epoch and the separation increase between epochs E2 and E3 relative to E1 in G23.389$+$00.185. The dashed lines mark the mean histogram values, while the dotted lines mark the median values.} 
\addtocounter{figure}{-1}
\end{figure}

\begin{figure}
\includegraphics[scale=0.52]{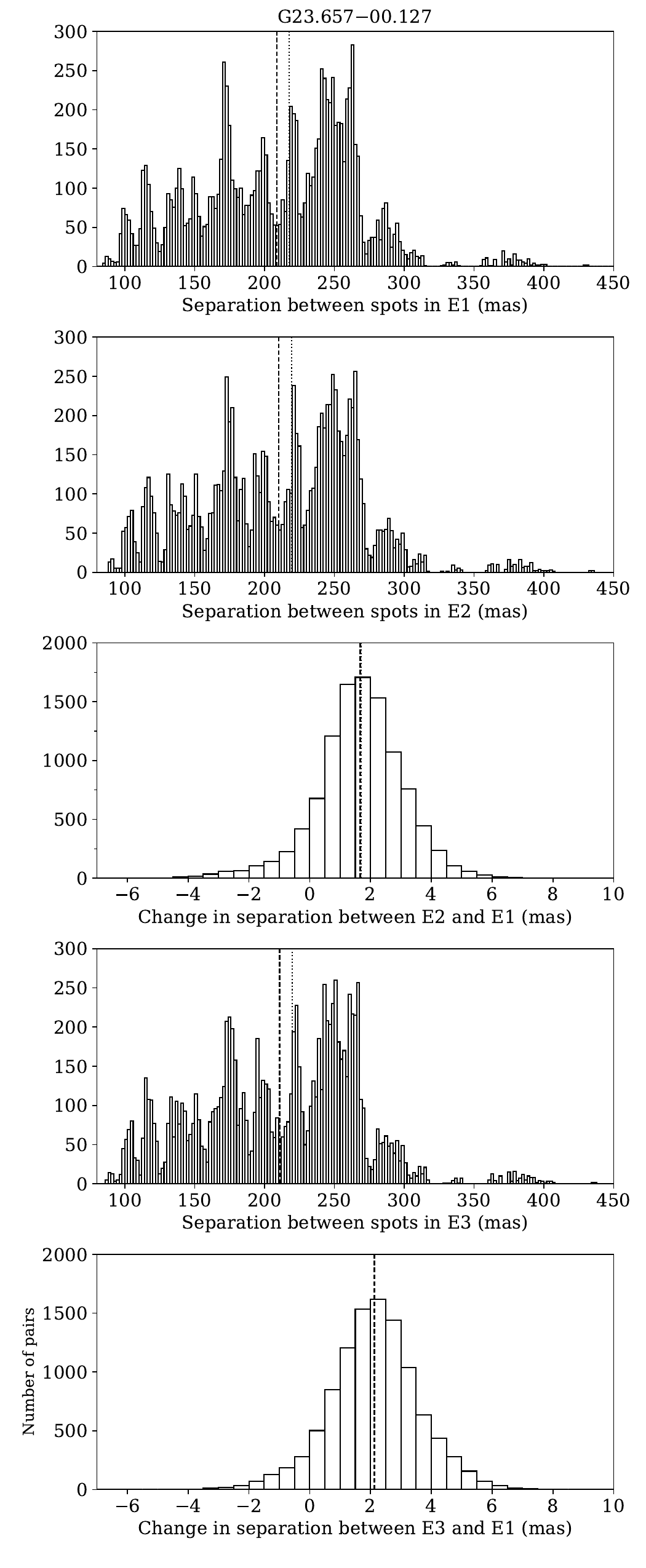}
\caption{Continued. Histograms of separations between northern and southern maser spot pairs in each epoch and the separation increase between epochs E2 and E3 relative to E1 in G23.657$-$00.127.} 
\label{histogramsapp}  
\end{figure}

\begin{figure}
\includegraphics[scale=0.52]{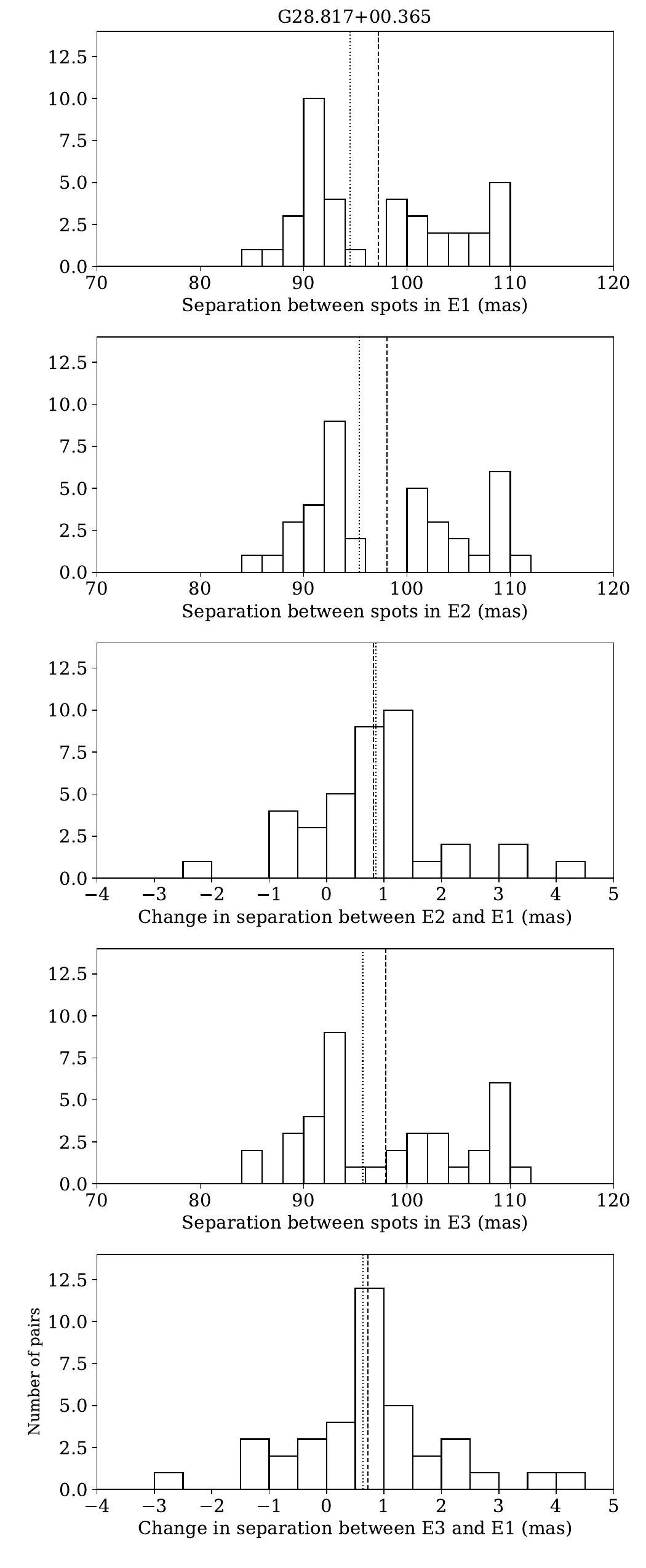}
\caption{Continued. Histograms of separations between eastern and western maser spot pairs in each epoch and the separation increase between epochs E2 and E3 relative to E1 in G28.817$+$00.365.} 
\label{histogramsapp}  
\end{figure}

\begin{figure}
\includegraphics[scale=0.52]{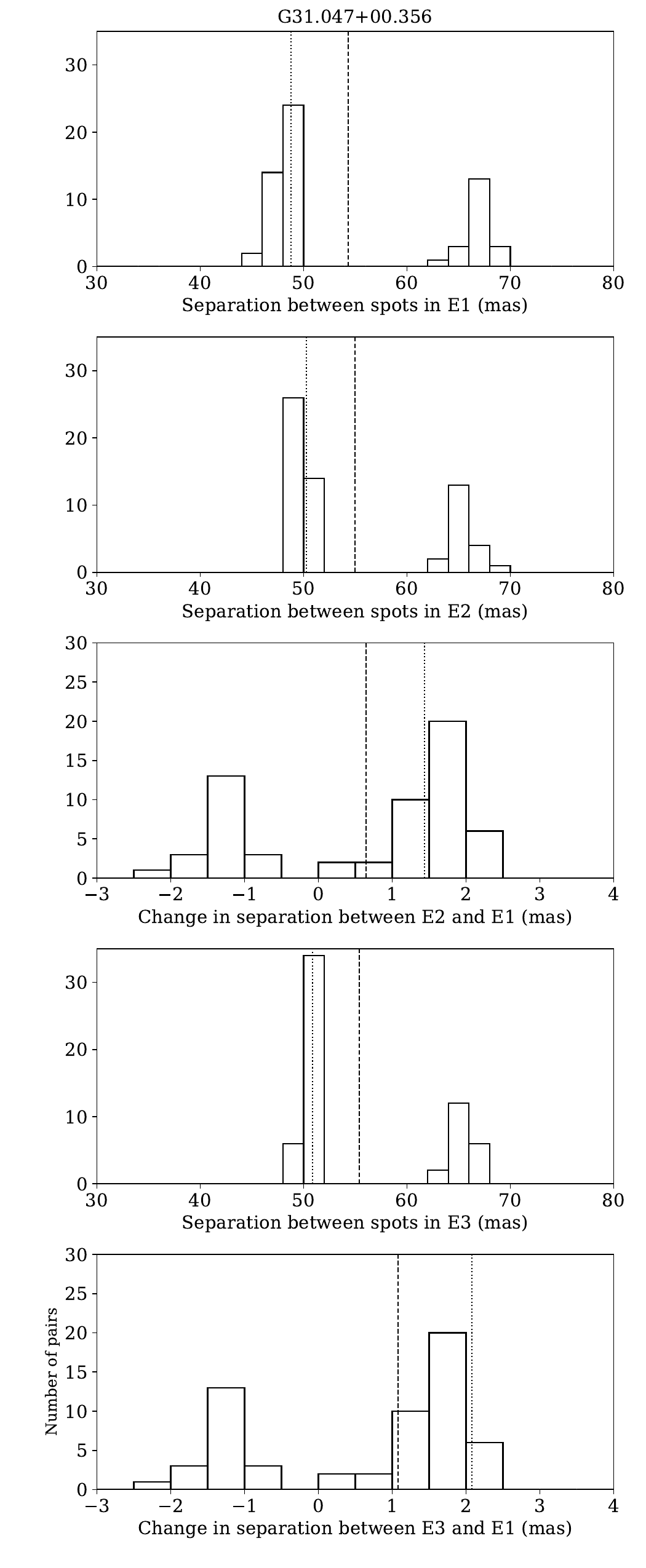}
\caption{Continued. Histograms of separations between eastern and western maser spot pairs in each epoch and the separation increase between epochs E2 and E3 relative to E1 in G31.047$+$00.356.} 
\label{histogramsapp}  
\end{figure}

\begin{figure}
\includegraphics[scale=0.52]{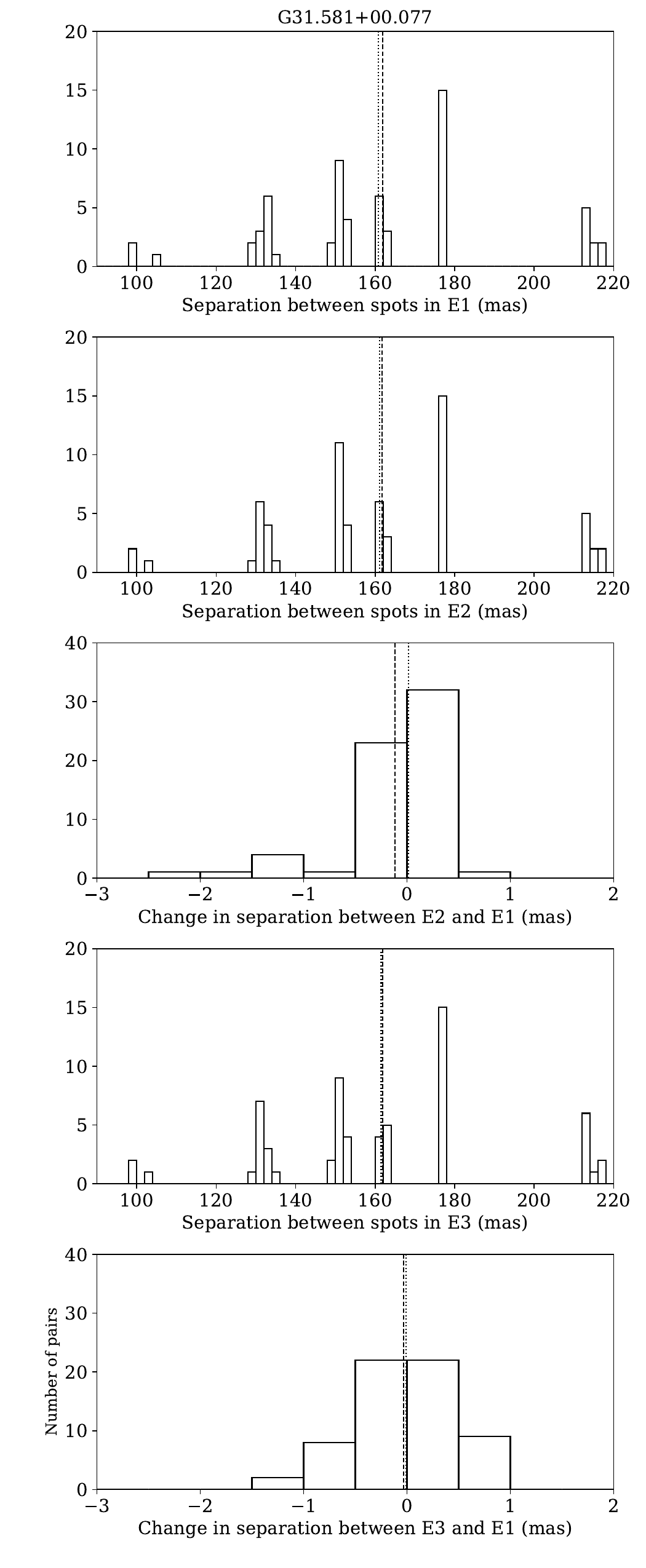}
\caption{Continued. Histograms of separations between eastern and western maser spot pairs in each epoch and the separation increase between epochs E2 and E3 relative to E1 in G31.581+00.077.} 
\label{histogramsapp}  
\end{figure}

\begin{figure*}
\centering
\includegraphics[width=0.49\textwidth]{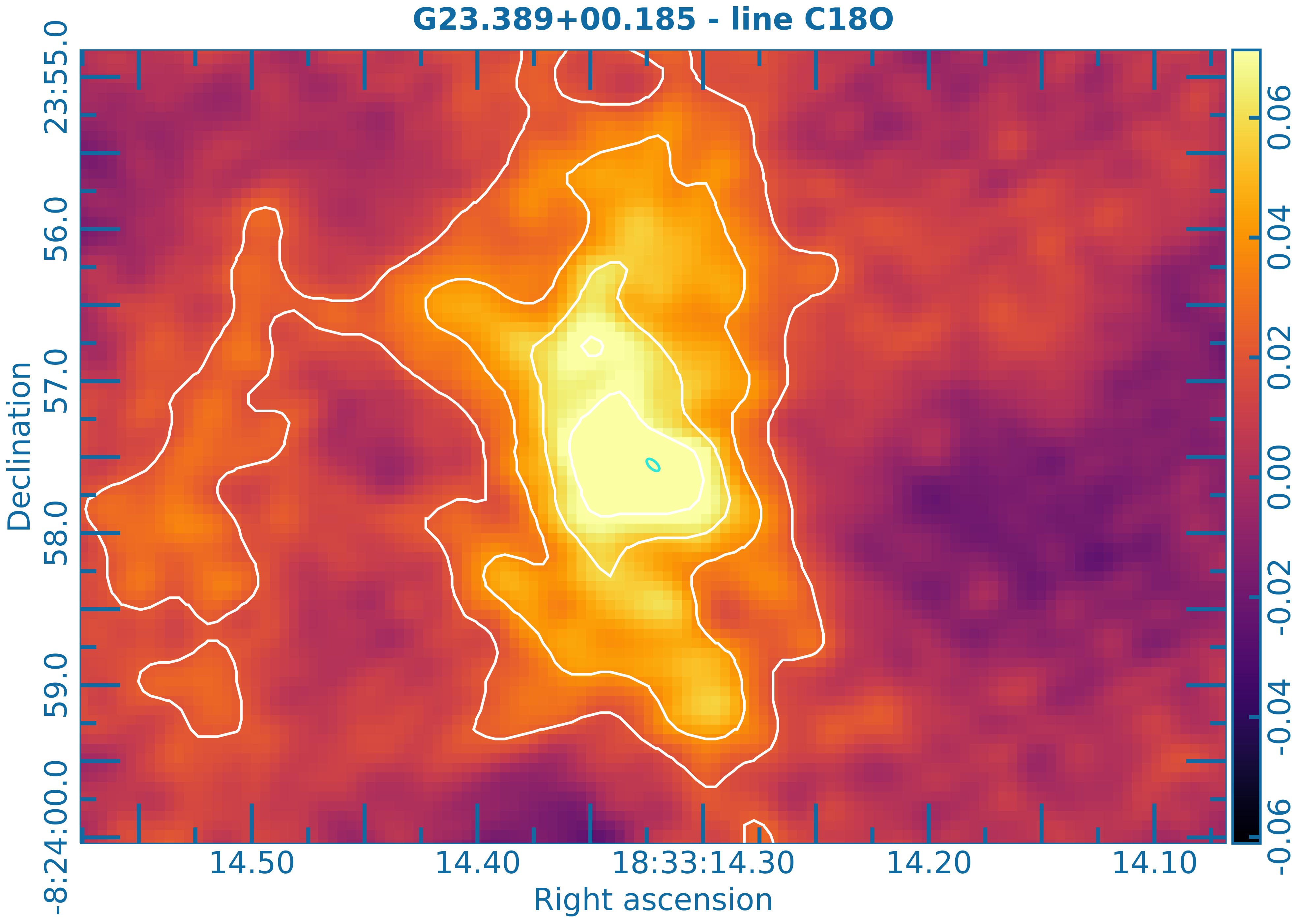}
\includegraphics[width=0.49\textwidth]{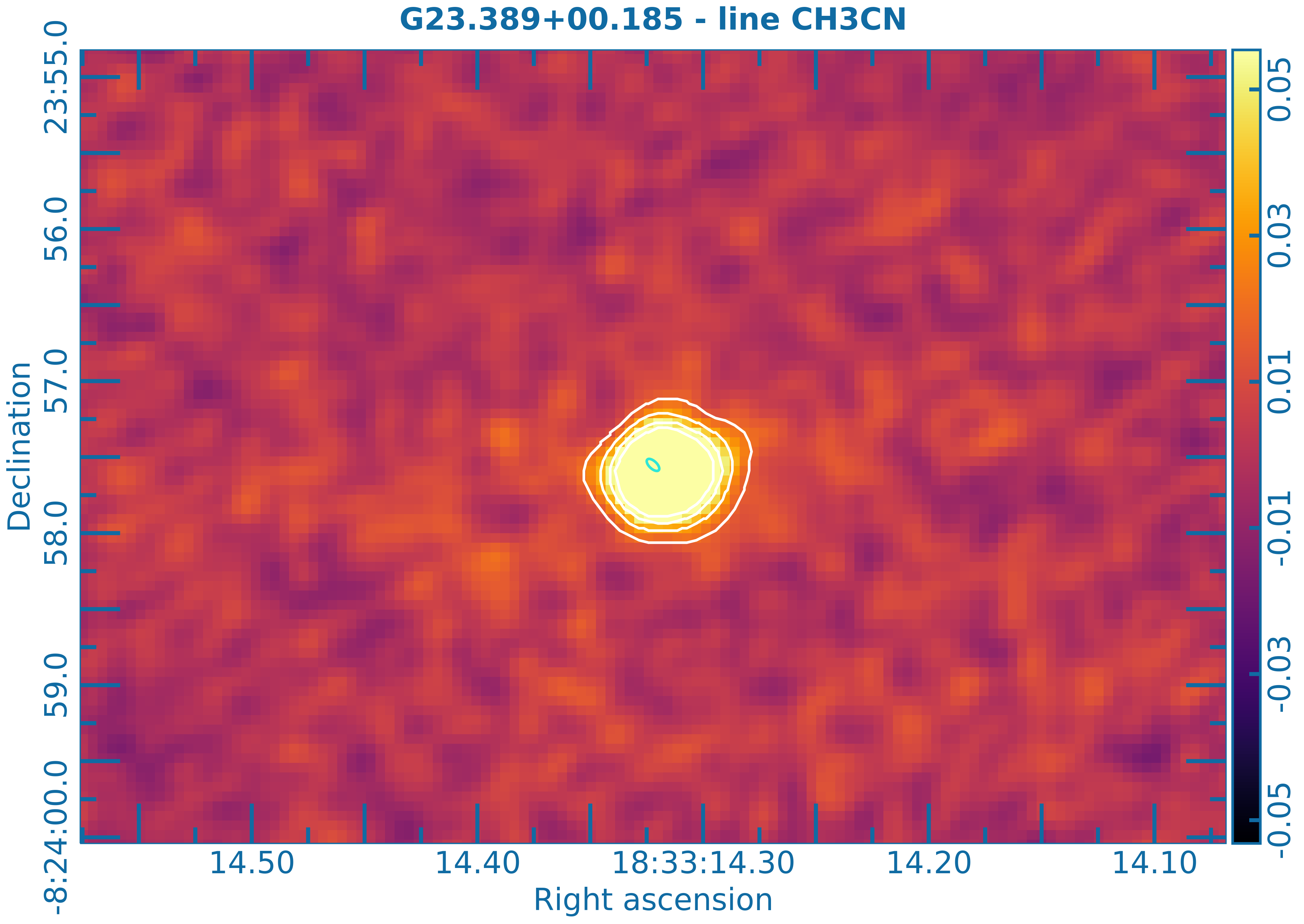}
\includegraphics[width=0.49\textwidth]{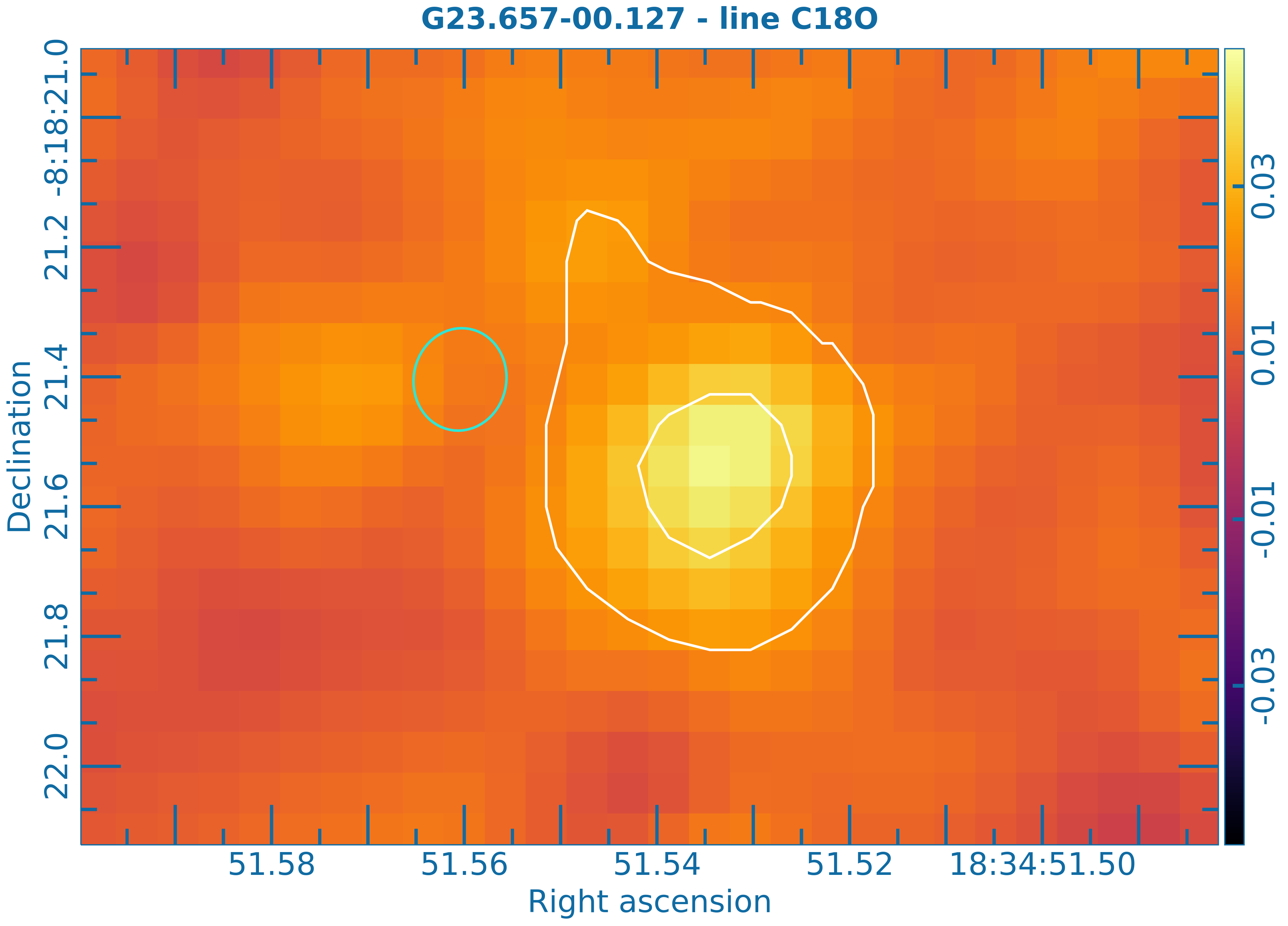}
\includegraphics[width=0.49\textwidth]{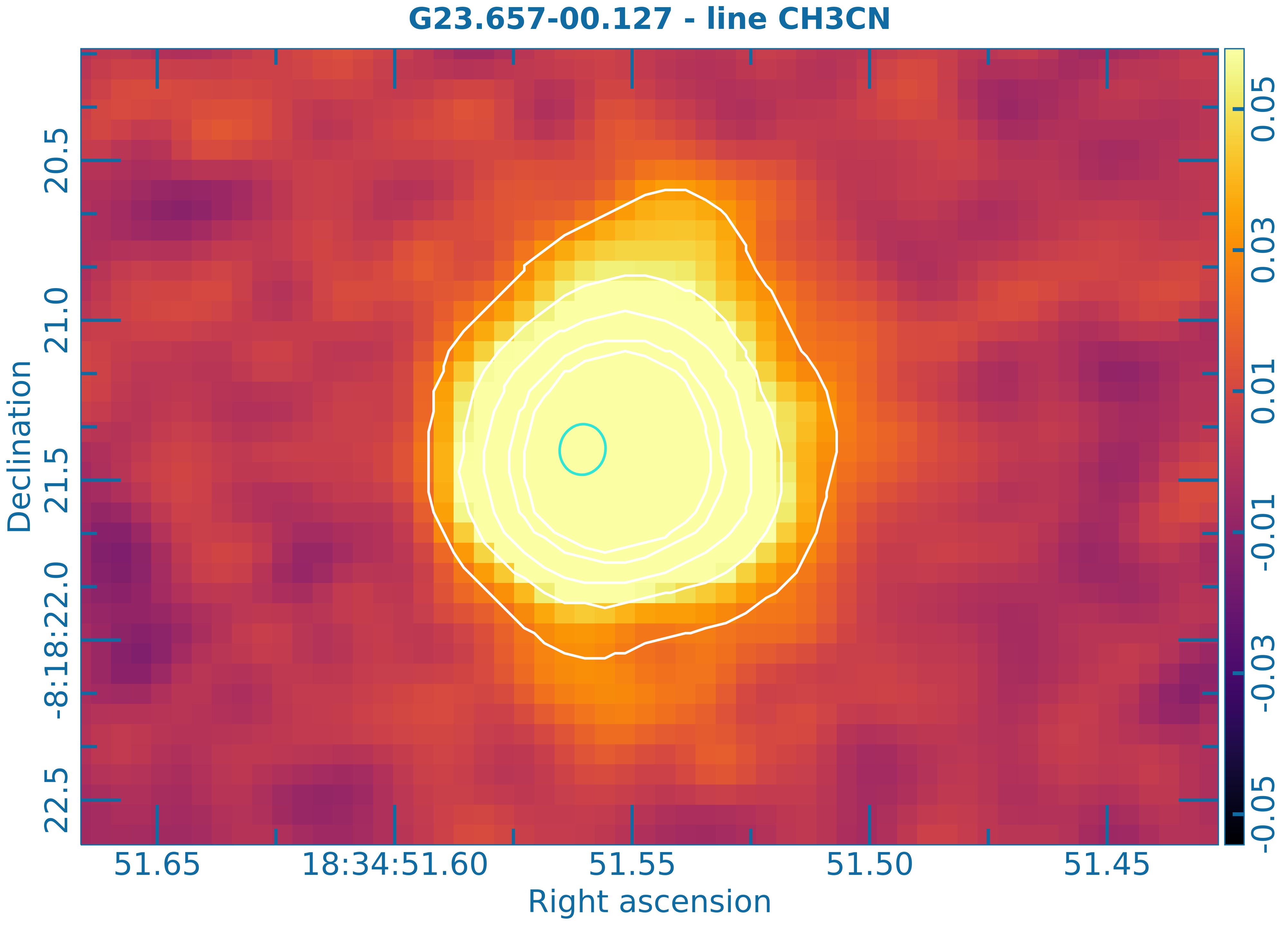}
\includegraphics[width=0.49\textwidth]{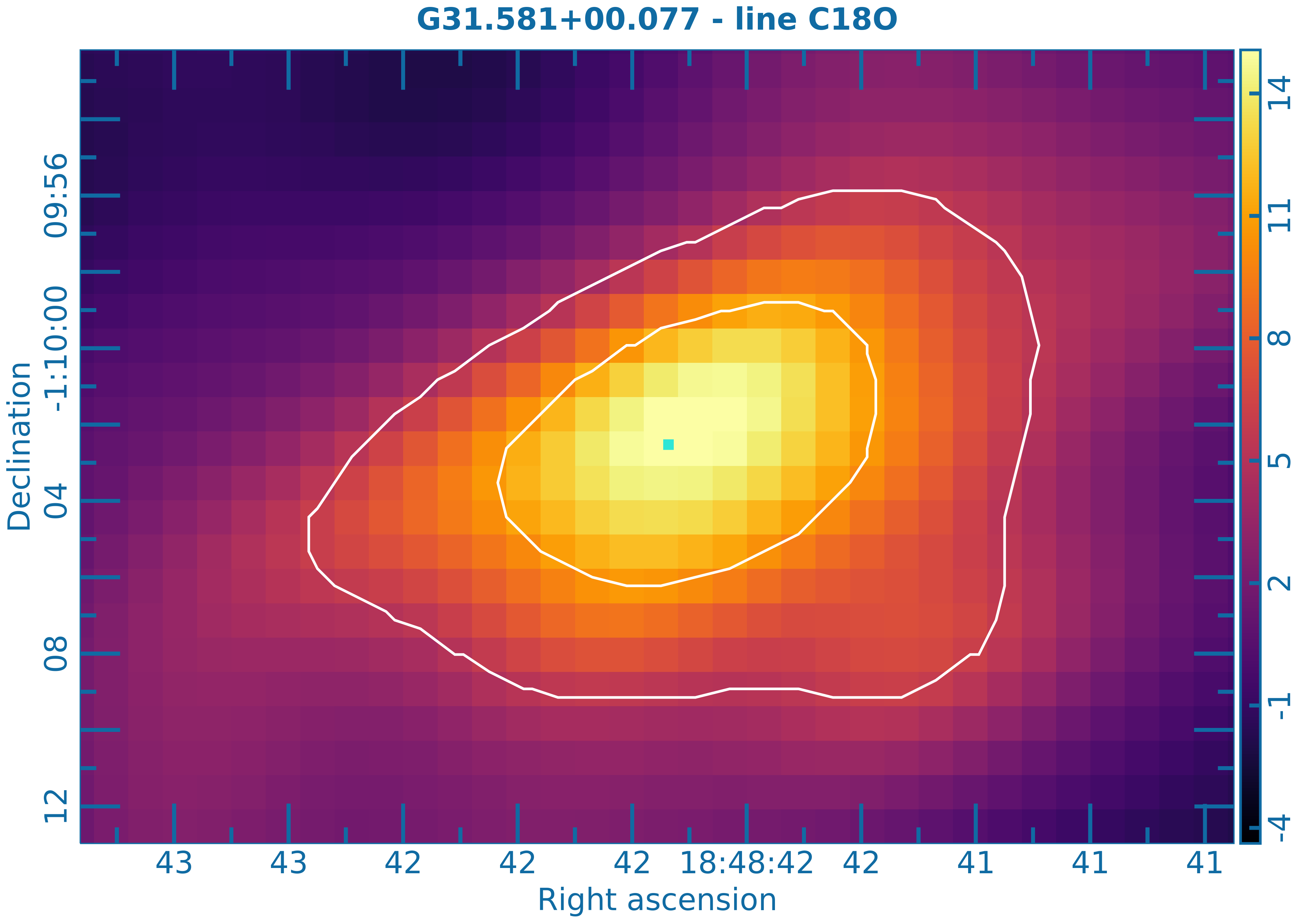}
\includegraphics[width=0.49\textwidth]{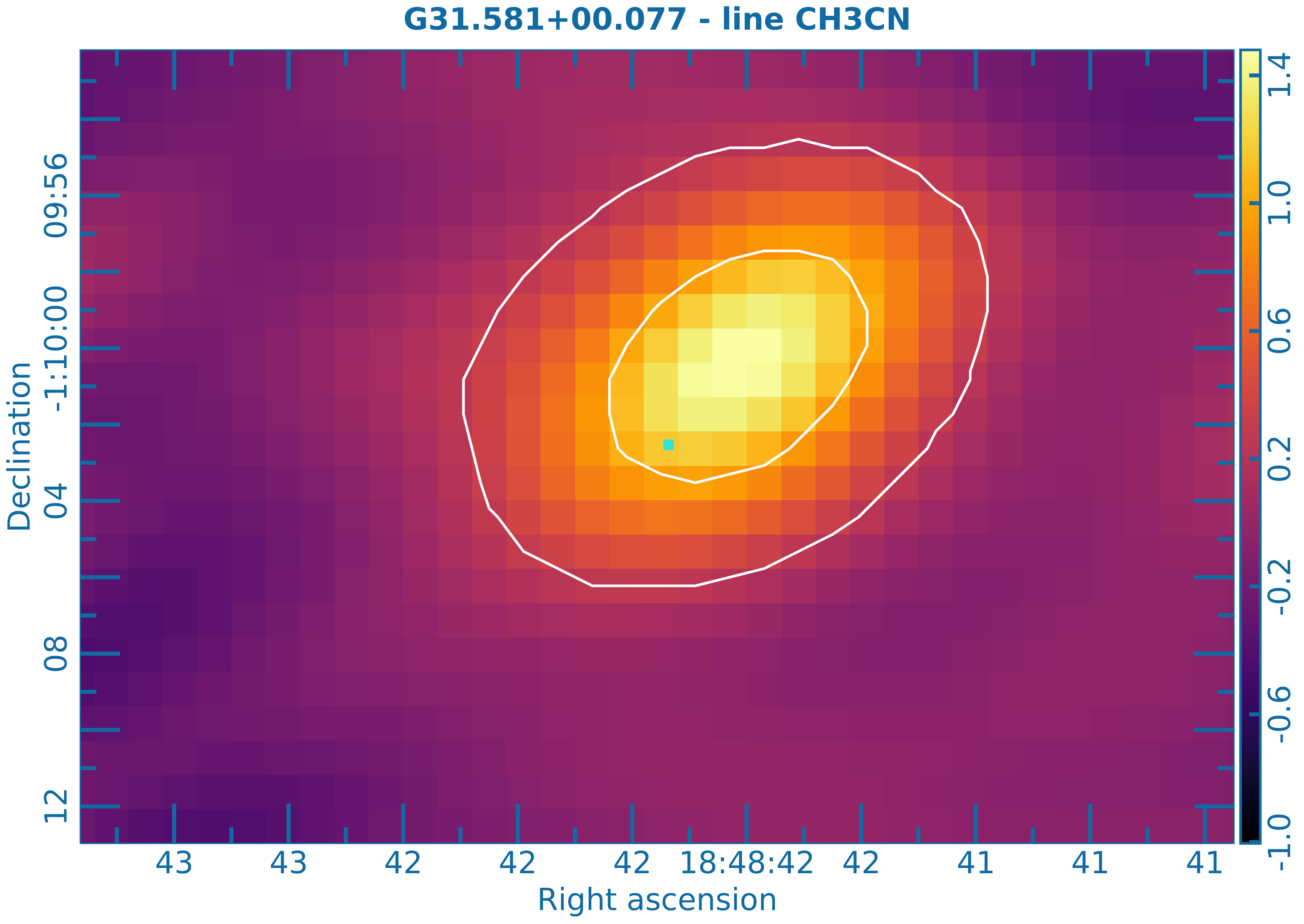}
\caption{Thermal emission closer to G23.389$+$00.185 (top panel), G23.657$-$00.127 (middle panel), and G31.581$+$00.077 (bottom panel) from the ALMA archive. Data were part of the large programme ALMAGal (ID: 2019.1.00195.L). The methanol emission is presented with the green ellipses. The colourful images correspond to C$^{18}$O (left) and CH$_3$CN (right) emission for channel maps taken at the line peak.}
\label{almacounter}
\end{figure*}

\onecolumn
\begin{longtable}{c r r c c c c c c c c}
\caption{\label{clouds} Parameters of 6.7~GHz methanol maser cloudlets used in the proper motion studies. The coordinates are first-epoch data and are relative to averaged proper motions of persistent cloudlets with linear motions; the same as in Figs~\ref{g23207cm}-\ref{g31581cm}. The (0,0) points are listed for each target. 
Cloudlets that are marked by the superscript $n$ were not used in the proper motion calculations due to their non-linear motions. 
The full-width at the half-maximum is given for the Gaussian fit of each cloudlet; $-$ denotes a non-Gaussian profile. In the case of G31.047$+$00.356, we list parameters for maser spots (see explanations in Sect.~3.1). Data for G23.657$-$00.127 are presented in more detail in Table A.1 in \cite{b20}}\\
\hline
 V$_{\rm p}$ & $\Delta$RA & $\Delta$Dec & $\mu_{\rm RA}$ & $\mu_{\rm Dec}$ & \multicolumn{3}{c}{Flux density} & \multicolumn{3}{c}{FWHM}\\
(km~s$^{-1}$) & (mas) & (mas) & (mas~yr$^{-1}$) & (mas~yr$^{-1}$) & E1 & E2 & E3 & E1 & E2 & E3\\
  & & & & &  \multicolumn{3}{c}{(Jy~beam$^{-1}$)} & \multicolumn{3}{c}{(km~s$^{-1}$)}\\
\hline                    
\endfirsthead
\caption{Continued.}\\
\hline
 V$_{\rm p}$ & $\Delta$RA & $\Delta$Dec & $\mu_{\rm RA}$ & $\mu_{\rm Dec}$ & \multicolumn{3}{c}{Flux density} & \multicolumn{3}{c}{FWHM}\\
(km~s$^{-1}$) & (mas) & (mas) & (mas~yr$^{-1}$) & (mas~yr$^{-1}$) & E1 & E2 & E3 & E1 & E2 & E3\\
  & & & & &  \multicolumn{3}{c}{(Jy~beam$^{-1}$)} & \multicolumn{3}{c}{(km~s$^{-1}$)} \\
\hline
\endhead
\hline
\endfoot
\hline
\endlastfoot
\multicolumn{4}{l}{\bf G23.207$-$00.377} & \multicolumn{7}{c}{\bf RA=18$^{\rm h}$34$^{\rm m}$55\fs20790, Dec=$-$08$^{\rm o}$49'11\farcs8421 (J2000)}\\
72.53 & -38.824 & +14.095 & -0.04 $\pm$ 0.01 & +0.19 $\pm$ 0.01 & 0.319 & 0.204 & 0.365 & 0.35 & 0.39 & 0.30\\ 
73.84 & +30.114 & -32.762 & +0.17 $\pm$ 0.01 & +0.09 $\pm$ 0.02 & 0.446 & 1.195 & 0.319 & 0.28 & 0.29 & 0.32\\ 
74.85 & +37.313 & -48.534 & +0.16 $\pm$ 0.01 & +0.04 $\pm$ 0.04 & 0.787 & 0.444 & 0.457 & 0.33 & 0.51 & 0.61\\
 & & & & & & & & 0.56 & 0.53\\ 
75.07 & +62.927 & -60.272 & +0.05 $\pm$ 0.03 & -0.10 $\pm$ 0.01 & 0.462 & 0.384 & 0.421 & 0.29 & 0.32 & 0.33\\ 
75.42 & -206.951 & -75.988 & -0.17 $\pm$ 0.01 & -0.00 $\pm$ 0.01 & 2.040 & 1.343 & 1.183 & 0.36 & 0.39 & 0.39\\ 
75.51 & +119.937 & -91.755 & +0.19 $\pm$ 0.01 & -0.15 $\pm$ 0.01 & 1.356 & 1.049 & 0.807 & 0.35 & 0.54 & 0.42 \\ 
75.69 & -158.935 & -191.421 & -0.21 $\pm$ 0.04 & -0.27 $\pm$ 0.05 & 0.130 & 0.307 & 0.371 & 0.32 & 0.29 & 0.29\\
76.30$^n$ & +50.295 & -54.298 & -0.13 $\pm$ 0.04 & -0.08 $\pm$ 0.01 & 10.212 & 2.264 & 1.914 & 0.40$^d$ & 0.36 & 0.35\\ 
76.66$^n$ & +52.242 & -43.572 & +0.54 $\pm$ 0.11 & +0.22 $\pm$ 0.03 &  & 3.528 & 3.978 & 0.40$^d$ & 0.47 & 0.44\\ 
77.14 & +62.746 & -49.179 & +0.06 $\pm$ 0.01 & -0.02 $\pm$ 0.01 & 18.302 & 5.211 & 4.615 & 0.36 & 0.50 & 0.65\\ 
77.49 & +92.042 & -46.048 & +0.05 $\pm$ 0.01 & -0.11 $\pm$ 0.01 & 7.326 & 3.939 & 4.621 & 0.45 & 0.55 & 0.37 \\
 & & & & & & & & & & 0.36\\ 
77.93 & +74.636 & -52.639 & -0.03 $\pm$ 0.03 & -0.07 $\pm$ 0.02 & 6.301 & 2.607 & 3.052 & 0.45 & 0.39 & 0.37\\
 & & & & & & & &  & 0.69 & 0.80\\ 
78.59 & +106.129 & -21.073 & -0.05 $\pm$ 0.01 & -0.15 $\pm$ 0.03 & 0.607 & 0.604 & 0.609 & $-$ & 0.41 & 0.42\\ 
78.59 & +116.001 & -17.315 & +0.09 $\pm$ 0.01 & -0.08 $\pm$ 0.03 & 0.949 & 0.370 & 0.317 & 0.54 & 0.40 & 0.49\\ 
79.20 & +113.382 & -18.202 & +0.05 $\pm$ 0.01 & -0.06 $\pm$ 0.01 & 2.835 & 1.765 & 1.608 & 0.31 & 0.19 & 0.30\\ 
79.38 & -73.128 & +141.111 & -0.06 $\pm$ 0.01 & +0.05 $\pm$ 0.01 & 0.709 & 0.268 & 0.340 & 0.23 & 0.32 & $>$1  \\
 & & & & & & & & &  & 0.21\\ 
79.86 & -94.615 & +184.008 & -0.08 $\pm$ 0.01 & +0.06 $\pm$ 0.04 & 0.260 & 0.298 & 0.329 & 0.22 & 0.33 & 0.31 \\ 
79.91 & +4.970 & +20.324 & +0.03 $\pm$ 0.01 & +0.11 $\pm$ 0.03 & 0.546 & 2.282 & 1.963 & 0.31 & 0.31 & 0.33\\ 
80.78$^n$ & -100.734 & +97.145 & -0.20 $\pm$ 0.05 & +0.18 $\pm$ 0.03 & 12.578 & 4.050 & 2.918  & 0.32 & 0.35 & 0.53\\
 & & & & & & & & 0.72 & 0.33\\ 
81.84 & -77.854 & +90.433 & -0.15 $\pm$ 0.04 & +0.20 $\pm$ 0.02 & 13.482 & 4.664 & 5.733 & 0.52 & 0.63 & 0.48\\ 
81.83 & -94.597 & +95.504 & +0.23 $\pm$ 0.01 & +0.17 $\pm$ 0.01 & 4.189 & 4.162 & 4.647 & 0.35 & $-$ &0.24 \\ 
82.76 & -89.469 & +96.954 & -0.47 $\pm$ 0.04 & +0.10 $\pm$ 0.01 & 2.309 & 9.441 & 14.526 & 0.41 & 0.37 & 0.38\\
  & & & & & & & &  &  & 0.57\\ 
84.87 & +14.185 & +62.763 & +0.17 $\pm$ 0.01 & -0.01 $\pm$ 0.01 & 1.584 & 1.920 & 2.485 & 0.34 & 0.40 & 0.40\\
 & & & & & & & & 0.42 &  &\\ 
\\
\multicolumn{4}{l}{\bf G23.389$+$00.185} & \multicolumn{7}{c}{\bf RA=18$^{\rm h}$33$^{\rm m}$14\fs32364, Dec=$-$08$^{\rm o}$23'57\farcs5461 (J2000)}\\
72.39 & -62.001 & +40.339 & -0.01 $\pm$ 0.01 & -0.03 $\pm$ 0.01 & 7.076 & 4.943 & 6.594 & 0.46 & 0.32 & 0.30\\ 
73.62 & +69.944 & +43.180 & +0.26 $\pm$ 0.06 & +0.65 $\pm$ 0.10 & 9.712 & 1.128 & 2.481 & 0.16 & 0.37 & 0.40\\ 
& && & & & & & 0.51\\
74.00 & +65.994 & +36.741 & +0.02 $\pm$ 0.02 & -0.16 $\pm$ 0.06 & 3.215 & 3.944 & 5.206 & 0.40& 0.25 & 0.27\\ 
& && & & & & & & 0.31\\
74.50 & -26.090 & +42.224 & +0.10 $\pm$ 0.01 & -0.08 $\pm$ 0.03 & 5.628 & 2.979 & 3.739 & 0.30 & 0.40 & 0.39\\ 
74.81 & +48.317 & +3.084 & -0.02 $\pm$ 0.01 & +0.00 $\pm$ 0.01 & 20.440 & 12.713 & 14.389 & 0.35 & 0.37 & 0.35\\ 
& && & & & & & 0.42& 0.33 & 0.40\\
75.34 & +16.794 & +73.460 & +0.00 $\pm$ 0.01 & +0.08 $\pm$ 0.01 & 23.521 & 14.063 & 16.308 & 0.24 & 0.28 & 0.28\\ 
& && & & & & & & & 0.37\\
76.04 & -75.817 & -81.790 & -0.12 $\pm$ 0.04 & -0.05 $\pm$ 0.03 & 4.859 & 5.303 & 7.542 & 0.16 & & \\ 
& && & & & & & 0.42 & 0.42 & 0.33\\
& && & & & & & 0.29 & 0.33 & 0.36\\
76.13 & -59.326 & -71.429 & -0.26 $\pm$ 0.08 & -0.26 $\pm$ 0.01 & 6.771 & 4.557 & 5.545 & 0.36 & 0.51 & 0.45\\ 
& && & & & & & & 0.14\\
77.01 & +37.558 & -1.593 & +0.04 $\pm$ 0.01 & -0.03 $\pm$ 0.02 & 0.693 & 0.338 & 0.396 & 0.30 & 0.31 & 0.30\\ 
77.45 & -15.369 & -84.216 & -0.01 $\pm$ 0.01 & -0.13 $\pm$ 0.01 & 0.533 & 0.216 & 0.306 & 0.23 & 0.27 & 0.26\\ 
\\
\multicolumn{4}{c}{\bf G28.817$+$00.365} & \multicolumn{7}{c}{\bf RA=18$^{\rm h}$42$^{\rm m}$37\fs34596, Dec=$-$03$^{\rm o}$29'40\farcs9556 (J2000)}\\
87.73 & -42.815 & -22.111 & -0.04 $\pm$ 0.04 & +0.07 $\pm$ 0.05 & 0.306 & 0.178 & 0.211 & $-$& 0.23 & 0.23 \\ 
89.75 & +28.677 & +27.623 & -0.06 $\pm$ 0.00 & -0.33 $\pm$ 0.07 & 0.660 & 0.264 & 0.278 &  0.68 & 0.37 & 0.64\\ 
 & & & & & & & & & & 0.25\\
90.01 & -60.296 & +14.278 & -0.01 $\pm$ 0.05 & -0.07 $\pm$ 0.02 & 0.124 & 0.085 & 0.089 & $-$& $-$ & 0.21\\ 
90.45$^n$ & +29.833 & +33.252 & +0.07 $\pm$ 0.01 & +0.10 $\pm$ 0.01 & 2.774 & 1.245 & 1.479 & 0.56 & 0.54 & 0.31 \\ 
91.33 & +33.482 & +52.819 & +0.03 $\pm$ 0.01 & +0.31 $\pm$ 0.08 & 2.309 & 1.360 & 1.365 & 0.32 & 0.22 & 0.26\\ 
 & & & & & & & & & 0.31 & 0.26\\
 & & & & & & & & & 0.32 & 0.32  \\
92.39$^n$ & +24.487 & +43.176 & -0.17 $\pm$ 0.14 & +0.57 $\pm$ 0.18 & 0.371 & 0.209 & 0.194 & 0.43 & 0.88 & 0.27\\ 
 & & & & & & & & 0.44 & & 0.40\\
92.65 & +27.411 & +60.646 & +0.09 $\pm$ 0.01 & +0.03 $\pm$ 0.03 & 1.924 & 1.129 & 1.292 & 0.38 & 0.40 & 0.42\\ 
\\
\multicolumn{4}{c}{{\bf G31.047$+$00.356} (spots)}& \multicolumn{7}{c}{\bf RA=18$^{\rm h}$46$^{\rm m}$43\fs85452, Dec=$-$01$^{\rm o}$30'54\farcs1557 (J2000)}\\
78.07 & +7.095 & +8.027 & -0.02 $\pm$ 0.01 & +0.02 $\pm$ 0.05 & 0.255 & 0.108 & 0.068 & \\ 
78.25 & +6.715 & +6.818 & -0.03 $\pm$ 0.01 & -0.01 $\pm$ 0.01 & 0.045 & 0.122 & 0.111 & $-$ &  0.46 & 0.38\\ 
78.42$^n$ & +7.633 & +5.233 & -0.20 $\pm$ 0.11 & +0.14 $\pm$ 0.07 & 0.122 & 0.067 & 0.114 & \\
78.60$^n$ & +6.881 & +4.722 & -0.11 $\pm$ 0.06 & +0.10 $\pm$ 0.05 & 0.076 & 0.082 & 0.073 & \\
78.78 & +6.638 & +4.113 & -0.01 $\pm$ 0.03 & +0.09 $\pm$ 0.01 & 0.300 & 0.228 & 0.174 & \\
78.95 & +6.758 & +3.839 & -0.00 $\pm$ 0.01 & +0.08 $\pm$ 0.01 & 0.610 & 0.540 & 0.492 & \\
79.13 & +6.625 & +3.621 & +0.06 $\pm$ 0.02 & +0.06 $\pm$ 0.01 & 0.981 & 0.826 & 0.996 &  0.71 & 0.69 & 0.71\\ 
79.30 & +6.593 & +3.584 & +0.07 $\pm$ 0.02 & +0.02 $\pm$ 0.01 & 0.943 & 0.756 & 0.930 & \\
79.48 & +6.807 & +3.180 & +0.03 $\pm$ 0.00 & +0.01 $\pm$ 0.02 & 0.641 & 0.562 & 0.689 & \\
79.65 & +7.274 & +3.109 & -0.01 $\pm$ 0.01 & -0.08 $\pm$ 0.05 & 0.433 & 0.463 & 0.597 & \\
79.83 & +7.893 & +2.172 & +0.00 $\pm$ 0.00 & -0.03 $\pm$ 0.03 & 0.278 & 0.604 & 0.641 & \\
80.00 & +8.222 & +2.092 & +0.02 $\pm$ 0.01 & -0.04 $\pm$ 0.02 & 0.382 & 1.090 & 1.040 &  0.34 & 0.60 & 0.54\\ 
80.18 & +8.406 & +2.072 & +0.03 $\pm$ 0.01 & -0.03 $\pm$ 0.03 & 0.471 & 1.657 & 1.880 &  \\
80.36 & +8.650 & +2.197 & -0.00 $\pm$ 0.02 & -0.08 $\pm$ 0.06 & 0.467 & 1.487 & 1.763 & \\
80.53 & +8.212 & +1.292 & +0.02 $\pm$ 0.00 & -0.03 $\pm$ 0.02 & 0.943 & 1.453 & 1.464 & \\
80.71 & +7.973 & +0.656 & +0.01 $\pm$ 0.01 & -0.04 $\pm$ 0.01 & 1.580 & 1.664 & 1.867 & 0.81 & 0.21 & 0.18\\ 
80.88 & +7.830 & -0.395 & +0.02 $\pm$ 0.01 & -0.03 $\pm$ 0.01 & 1.351 & 1.416 & 1.584 &  & 0.65 & 0.61\\ 
81.06 & +7.427 & -0.946 & +0.04 $\pm$ 0.01 & -0.05 $\pm$ 0.01 & 1.142 & 1.274 & 1.484 & \\
81.06 & -51.775 & -28.434 & +0.17 $\pm$ 0.02 & +0.07 $\pm$ 0.05 & 0.198 & 0.232 & 0.357 & $-$ & 0.29 & 0.29\\ 
81.23 & +6.718 & -1.695 & +0.06 $\pm$ 0.01 & -0.05 $\pm$ 0.01 & 0.920 & 0.903 & 0.965 & \\
81.41 & +5.790 & -2.762 & +0.12 $\pm$ 0.01 & +0.02 $\pm$ 0.01 & 0.357 & 0.393 & 0.393 & \\
82.82 & -39.990 & -6.322 & -0.28 $\pm$ 0.01 & +0.07 $\pm$ 0.01 & 0.115 & 2.263 & 3.496 &\\
82.99 & -39.859 & -6.220 & -0.30 $\pm$ 0.01 & +0.04 $\pm$ 0.01 & 0.528 & 1.019 & 0.560 & 0.28 & 0.29 & 0.31\\ 
\\
\multicolumn{4}{c}{{\bf G31.581$+$00.077}} & \multicolumn{7}{c}{\bf RA=18$^{\rm h}$48$^{\rm m}$41.94582, Dec=$-$01$^{\rm o}$10'2\farcs4794 (J2000)}\\
91.60$^n$ & +34.405 & -128.074 & -0.00 $\pm$ 0.11 & +0.21 $\pm$ 0.06 & 0.299 & 0.102 & 0.110 & 0.31 & 0.34 & 0.37\\ 
95.64 & -71.099 & -48.651 & -0.01 $\pm$ 0.00 & -0.04 $\pm$ 0.00 & 2.630 & 1.083 & 1.822 & 0.33 & 0.34 & 0.33 \\ 
97.75 & -132.089 & +8.874 & -0.01 $\pm$ 0.00 & +0.06 $\pm$ 0.00 & 0.266 & 0.132 & 0.211 & $-$ & 0.24 & 0.22 \\ 
98.01$^n$ & -33.952 & -55.679 & -0.01 $\pm$ 0.06 & +0.04 $\pm$ 0.02 & 0.741 & 0.202 & 0.237 & 0.29 & 0.93$^{*}$ & 0.30 \\ 
98.45 & -38.817 & -52.326 & +0.07 $\pm$ 0.04 & -0.10 $\pm$ 0.05 & 0.319 & 0.141 & 0.247 & 0.36 & 0.93$^{*}$ & 0.31\\ 
98.80 & +78.855 & +45.003 & -0.04 $\pm$ 0.01 & +0.02 $\pm$ 0.03 & 1.628 & 1.078 & 2.020 & 0.30 & 0.39 & 0.33 \\ 
99.24 & +82.712 & +39.239 & -0.04 $\pm$ 0.03 & +0.07 $\pm$ 0.06 & 0.847 & 0.367 & 0.737 & 0.30 & 0.32 & 0.31\\ 
99.68 & +80.440 & +7.862 & +0.03 $\pm$ 0.03 & -0.01 $\pm$ 0.02 & 0.749 & 0.517 & 1.081 & 0.37 & 0.45 & 0.40\\ 
\\
\hline
\end{longtable}
\tablefoot{$^d$ in G23.207$-$00.377 denotes a case when a single Gaussian profile in the first epoch (E1) becomes a double in E2 and E3.}

\end{appendix}
\end{document}